\documentclass [12pt]{article}
\usepackage{amssymb}
\usepackage{amsmath}
\usepackage{graphicx}
\usepackage{amscd}
\usepackage {longtable}
\usepackage{epsfig}
\date{}

\title{QUADRUPOLE OSCILLATIONS AS PARADIGM OF THE CHAOTIC MOTION IN NUCLEI.(Part 1)}

\author{V.P.Berezovoj\footnote{e-mail: berezovoj@kipt.kharkov.ua},
 Yu.L. Bolotin, V. Yu. Gonchar,
${}^{\star}$M.Ya.Granovsky,\\ V.N. Tarasov}

\begin{document}

\maketitle

\begin{center}{\it
National Science Center "Kharkov Institute of Physics and Technology", Kharkov, 310108, Ukraine\\

${}^\star$Pro Training Tutorial Institute, 18 Stoke Av, Kew 3101,
Melbourne, Vic., Australia }

\vspace{1cm}
\end{center}

We have presented a complete description of classical dynamics
generated by the Hamiltonian of quadrupole nuclear oscillations
and identified those peculiarities of quantum dynamics that can be
interpreted as quantum manifestations of classical stochasticity.
Particular attention has been given to investigation of classical
dynamics in the potential energy surface with a few local minima.
A new technique is suggested for determination of the critical
energy of the transition to chaos. It is simpler than criteria of
transition to chaos connected with one or another version of
overlap resonances criterion. We have numerically demonstrated
that for potential with a localized unstable region motion becomes
regular at the high energy again, i.e. transition
regularity-chaos-regularity $\left( {R - C - R} \right)$ takes
place for these potentials. The variations of statistical
properties of energy spectrum in the process of $R - C - R$
transition have been studied in detail. We proved that the type of
the classical motion is correlated with the structure of the
eigenfunctions of highly excited states in the $R - C - R$
transition. Shell structure destruction induced by the increase of
nonintegrable perturbation was analyzed.

\vspace{2cm}


\newpage

\tableofcontents

\section{Introduction}

\subsection{Formulation of the problem}

Quite often in the different divisions of physics statistical
properties are introduced by means of postulates or hypotheses
that are self-justified within some limiting case. Such an
approach suggests an existence of a priori mechanism of randomness
the nature of which lies usually outside the theory under
discussion. For systems with a small number of degrees of freedom
the results obtained in the thermodynamic limit look at least
questionable. At the moment, however, one can consider as
rigorously established fact an existence of such dynamical systems
with a small number of degrees of freedom for which under certain
conditions classical motion is distinguishable from random no
matter what the definition of this notion we are using \cite{1, 2,
3, 4, 5, 6}. Typical features of these systems are nonlinearity
and absence of both external source of randomness and dissipation
the mechanism of which is equivalent to the existence of random
forces on a molecular level. Thus, using such synonyms for the
term "random" as "chaotic", "stochastic", "irregular" one can
state that there exist such systems with a finite number of
degrees of freedom, for which these notions express adequately
internal fundamental properties that comprise an important and
interesting subject for investigation. For the last 30-40 years
has been taken place a rather difficult recognition of the fact
that chaotic motion is as usual in systems with more than one
degree of freedom as regular motion. Instances of a chaotic motion
have been found practically in every field of physics, and their
number permanently  continues to grow \cite{1, 2, 3, 4, 5, 6}.

Recent progress in understanding of the chaotic aspects of
nonlinear dynamical systems is causing a rebirth of interest in
the nuclear dynamics. Indeed, because of the richness of
experimental data and sufficiently  precision of the theory, the
nuclear dynamics provides a very  useful realistic model for
studying classical dynamical chaos and quantum manifestations of
the classical stochasticity.

Conception of chaos has been introduced in the nuclear theory
within the last fifteen  years \cite{7, 8, 9, 10, 11, 12, 13, 14,
15, 16, 17, 18, 19, 20}. This conception brought birth to the new
notion in the nuclear structure \cite{7,10,11,13,21}, nuclear
reactions \cite{8,14,20}, could resolve a sequence of the very old
contradictions in the nuclear theory \cite{13,22}. A radically new
universal approach to the problem of statistical properties of the
energy spectra was developed on the basis of the general nonlinear
theory of dynamical systems \cite{23}. Considerable advances have
been made in the area of concrete nuclear effects: single particle
in a deformed potential \cite{11}, nuclear fission \cite{9},
Ericson's fluctuations \cite{24}, dynamics of the $3\alpha $ -
linear chain \cite{25}, transition order-chaos-order in the
roto-vibrational nuclear model \cite{111,112} etc. Finally,
straightforward observations of the chaotic regimes in the course
of mathematical simulations of the heavy-ions reactions \cite{8}
are evidence in favour the general considerations.

According to Baranger \cite{13} there are two possible
philosophies in nuclear physics.

Philosophy I. Nuclei are complicated, and chaos comes out of this
complication. We expect to find chaos almost everywhere in nuclear
physics. The interesting information is contained in those few
places in which chaos is absent. We must look for non-chaos.

Philosophy II. Chaos is a property of simple systems; otherwise
it's no fun at all. The interesting new information is to be found
in those simple areas of nuclear physics which we used to think we
could understand, but which turn out to be chaotic. We must look
for chaos.

The basis of present report is the philosophy of simple chaos -
Philosophy II. Within the limits of this philosophy a general
investigation of any nonlinear dynamical system involves the
following steps.

1. An investigation of classical phase space with the aim of
detection of chaotic regimes; numerical investigation of the
classical equations of motion .

2. Analytical estimation of the critical energy  for the onset of
global stochasticity.

3. Test for quantum manifestations of classical stochasticity
(QMCS) in the energy spectra, eigenfunctions and wave packet
dynamics.

 4. A consideration of the interrelationship between
stochastic dynamics and concrete physical effects.

The basic subject matter of proposing report is to realize the
outlined program, at least partially, for the large-amplitude
quadrupole oscillations of nuclei (QON). We shall organize this
paper as follows.

In the first part (section 2.1-2.7) we present complete
description of classical dynamics generated by the Hamiltonian of
quadrupole oscillations. This description contains

1. The scale analysis of the Hamiltonian (QON).

2. The discussion of the topological peculiarities of deformation
potential.

3. The analytical estimations of critical energy of the transition
to chaos.

4. The construction of the approximate integrals of motion with
the help of normal forms.

5. The analysis of dynamics in the region of parameters space
where the deformation potential has some local minima; the
introduction of the concept of mixed state.

6. The general conception of reconstruction of regular motion at
high energies for the systems with localized region of
instability; the transition regularity-chaos-regularity as a
particular realization of this conception.

7. Demonstration of the chaotic regimes in heavy-ions reactions.

In the second part 
peculiarities of quantum dynamics of considered system, which can
be interpret as the QMCS. For this purpose in this part of the
report were discussed

1. The Birkhoff's-Gustavson's quantum normal forms.

2. The change of statistical properties of the energy spectra in
$R - C - R$ transition.

3. The change of the structure of wave functions in $R - C - R$
transition.

4. The destruction and the reconstruction of the shell structure
in $R - C - R$ transition.

5. The dynamics of wave packets.

Thus, the distinctive feature of the suggested report is that all
complex of questions connected with the investigation both
classical stochasticity and QMCS is considered in the context of
the unique dynamical system.

As for circle of readers whom this report is counted on, then on
the one hand we would like to pay attention of experts in all
areas of chaos to perspective field of application of general
theory of nonlinear dynamical systems, and on the other hand to
persuade nuclear experts that ideology  of simple chaos can be
equally useful just as traditional statistical approaches.

For convenience of the latter a brief summary of definitions and
concepts used in description of stochastic motion in Hamiltonian
systems will be given in section 1.2.

\subsection{Classical stochasticity: some definitions}

The classical dynamics of Hamiltonian systems with N degrees of
freedom is described by the canonical equations of motion
\begin{equation}\label{1.2.1}
  \dot p_i  =  - \frac{{\partial H}}
{{\partial q_i }},\quad \dot q_i  = \frac{{\partial H}} {{\partial
p_i }}\quad \left( {1 \leqslant i \leqslant N} \right)
\end{equation}
where $H\left( {\vec p,\vec q} \right)$ is the Hamiltonian of
system. The function of dynamical variables $F\left( {\vec p,\vec
q} \right)$ is such that
\begin{equation}\label{1.2.2}
  \left\{ {F,H} \right\} = 0
\end{equation}
where $\left\{ {.,.} \right\}$ are Poisson's brackets and it is
called the integral of motion. If there are $N$ independent
integrals $ F_i \left( {1 \leqslant i \leqslant N} \right) $, such
that $ \left\{ {F_i ,F_j } \right\} = 0 $, then the system
(\ref{1.2.1}) is integrable.

For integrable system one can introduce such canonical variables
action-angle $\left( {I_i ,\theta _i } \right)$ that  Hamiltonian
will be the function of only variables of action
\begin{equation}\label{1.2.3}
  H\left( {\vec p,\vec q} \right) = H\left( I \right)
\end{equation}

The region labeled by fixed $I$ , to which the motion is confined,
is therefore an $N$ -dimensional torus in the $2N$ -dimensional
phase space. Any phase point that lies on a given torus at any
time remains on it for all future times, so the torus itself is
invariant under the Hamiltonian flow, and is known as an invariant
torus. The motion on an invariant torus is given by

\begin{equation}\label{1.2.4}
  \theta _i \left( t \right) = \omega _i t + \theta _i \left( 0 \right)
\end{equation}
where angular frequency vector $\omega _j $ can be written as
\begin{equation}\label{1.2.5}
  \omega _i  = \frac{{\partial H}}
{{\partial I_i }}
\end{equation}
In this case the motion is a quasiperiodic function of time with
at least $m$ independent frequencies. If the $\omega _k $ are not
rationally related, that is, there are no integers $s_i $ , such
that
\begin{equation}\label{1.2.6}
  \sum\limits_i {s_i \omega _i  = 0\left( {s \ne 0} \right)}
\end{equation}
then the phase point passes arbitrary close to every point of the
torus and it is not difficult to show that the time average of
function is equal to the average of the function over the angle
variables. Thus the motion on the torus is ergodic. If
(\ref{1.2.6}) is satisfied for some nonzero integer vector s, than
is a resonance, and $N$ -tori are made up of tori of lower
dimension, to which the motion is confined. Such $N$ -tori are
degenerated. The particular case of integrable system is the
system with separable variables.

In the general case the system with two or more degrees of freedom
is not integrable and can perform both quasiperiodic (regular) and
stochastic motion. The distinctive feature of stochastic motion is
instability, exhibiting in exponential divergence of close
trajectories. If $x(t)$ and $x'(t)$ are two trajectories close at
$t = 0$ in phase space, then during stochastic motion
\begin{equation}\label{1.2.7}
  \Delta \left( t \right) = \left| {x\left( t \right) - x'\left( t \right)} \right| \approx \Delta \left( 0 \right)e^{\sigma _i \left( t \right)}
\end{equation}
at sufficiently small $\Delta \left( 0 \right)$. Quantity $\sigma
_i $ is called the maximal Liapunov exponent and is determined as
\begin{equation}\label{1.2.8}
  \sigma _i \left( x \right) = \lim _{t \rightarrow \infty, \Delta (0)\rightarrow 0}  \left( {\frac{1}
{t}\ln \frac{{\Delta \left( t \right)}} {{\Delta \left( 0
\right)}}} \right)\left( {x = x\left( 0 \right)} \right)
\end{equation}
The system with $D$ -dimensional phase space is characterized by
the set of $\Delta $ Lyapunov exponents $\sigma _i $ , which are
numbered in order of decreasing. The points belonging to
trajectory possess equal values $\sigma _i \left( x \right)$ . The
motion is called stochastic if for trajectory $\sigma _i  > 0$ and
regular if $\sigma _i  = 0$ .

Kolmogorov's entropy $h$ is connected with Lyapunov exponents. For
given stochastic trajectory
\begin{equation}\label{1.2.9}
  h = \sum\limits_{\sigma _i  > 0} {\sigma _i }
\end{equation}

At stochastic motion the correlation function
\begin{equation}\label{1.2.10}
B\left( {f,g,\tau } \right) = \left\langle {f\left( {t + \tau }
\right)g\left( t \right)} \right\rangle  - \left\langle {f\left( t
\right)} \right\rangle \left\langle {g\left( t \right)}
\right\rangle
\end{equation}
(corner brackets $\left\langle {...} \right\rangle $ denote the
averaging on time, and $ f\left( t \right) = f\left( {\vec p,\vec
q} \right)$ and $ g\left( t \right) = g\left( {\vec p,\vec q}
\right) $ tends to zero with the growth of $t$
\begin{equation}\label{1.2.11}
 \lim _{\tau  \to \infty } B\left( {f,g;\tau } \right)
= 0
\end{equation}
This property is called the mixing.

Power spectrum $S(\omega )$ of the dynamic value $f\left( t
\right)$ is determined by the expression
\begin{equation}\label{1.2.12}
S\left( {f;\omega } \right) =
 \frac{1}{{2\pi }}\int\limits_{ -
\infty }^\infty  {B\left( {f,f;\tau } \right)e^{ - i\omega \tau }
d\tau }
\end{equation}
For the motion with mixing the power spectrum $S(\omega )$ is
continuous, and for the regular motion it is discrete:
\begin{equation}\label{1.2.13}
  S\left( {f,\omega } \right) = \sum\limits_k {A_k \delta \left( {\omega  - \omega _k } \right)}
\end{equation}
here $A_k $ form the countable sequence, and frequencies $\omega
_k $ are the combinations of frequencies $\omega _j $ of
quasiperiodic motion.

\section{Regular and chaotic dynamics of \\the quadrupole
oscillations}

\subsection{ Hamiltonian}

It can be shown \cite{26} that using only the transformation
properties of the interaction, the deformational potential of
surface quadrupole oscillations of nuclei takes on the form
\begin{equation}\label{2.1.1}
 U\left( {a_0 ,a_2 } \right) = \sum\limits_{m,n} {C_{mn}
  \left( {a_0^2  + 2a_2^2 } \right)^m a_0^n } \left( {6a_2^2  - a_0^2 } \right)^n
\end{equation}
where $a_0 $ and $a_2 $ are internal coordinates of nuclear
surface at quadrupole oscillations
\begin{equation}\label{2.1.2}
R \left( {\theta ,\varphi } \right) = R_0 \left\{ {1 + a_0 Y_{2,0}
\left( {\theta ,\varphi } \right) + a_2 \left[ {Y_{2,2} \left(
{\theta ,\varphi } \right) + Y_{2, - 2} \left( {\theta ,\varphi }
\right)} \right]} \right\}
\end{equation}
Constants $C_{mn}$ can be considered as phenomenological
parameters, which within the limits of the particular models or
approximations (for instance, ATDHF theory) can be directly
related to the effective interaction of the nucleons in nucleus
\cite{27}. Considering that at the construction of (\ref{2.1.1})
only transformation properties of interaction are used, then this
expression describes potential energy of quadrupole oscillations
of a liquid drop of any nature, including only the specific
character of a drop in the coefficients of  expansion $C_{mn}$ .

Restricting ourselves with the members of the fourth degree in the
deformation, and assuming the equality of mass parameters for two
independent directions, we get the following $C_{3\nu } $
 - symmetric Hamiltonian
\begin{equation}\label{2.1.3}
H = \frac{{p_x^2  + p_y^2 }} {{2m}} + U\left( {x,y;a,b,c} \right),
\end{equation}
where
\[ U\left( {x,y;a,b,c} \right) = \frac{a}
{2}\left( {x^2  + y^2 } \right) + b\left( {x^2 y - \frac{1} {3}y^3
} \right) + c\left( {x^2  + y^2 } \right)^2,\]
\[  x = \sqrt 2 a_{2,} \quad y = a_0\]
\begin{equation}\label{2.1.3a}
  {2} = C_{10} ,\quad b = 3C_{01} ,\quad c = C_{20}
\end{equation}

The potential $ U\left( {x,y;a,b,c} \right)$ is the generalization
of the well-known Henon - Heiles potential \cite{28} with the one
important difference that the motion in the potential
(\ref{2.1.3a}) is finite for all energies. This is particularly
important for the quantum case, where the potential (\ref{2.1.3a})
ensures the existence of stationary states (instead of
quasistationary ones as with the Henon-Heiles potential).

$C_{3\nu } $ - symmetry of  potential surface becomes obvious in
polar coordinates
\begin{equation}\label{2.1.4}
 x = \beta \sin \gamma ,y = \beta \cos \gamma
\end{equation}
where $\beta $ is the so-called parameter of deformation of axial
symmetric nucleus, and $\gamma $ is the parameter of nonaxiality.

In these coordinates
\begin{equation}\label{2.1.5} U\left( {\beta
,\gamma ;a,b,c} \right) = \frac{1} {2}a\beta ^2  - \frac{1}
{3}b\beta ^3 \cos 3\gamma  + c\beta ^4
\end{equation}
There are three possibilities of introduction of the typical unit
of length $l_{0i} $ $(i = 1,2,3)$ for the Hamiltonian
(\ref{2.1.3})

    1) as the distance $l_{01} $, at which the contributions from harmonic
    and cubic terms become comparable

    2) as the distance $l_{02} $
, at which the contributions from harmonic and biquadratic terms
become comparable

    3) as the distance $l_{03} $
, at which the contributions from cubic and biquadratic terms
become comparable.

Scaling of the principal physical values (scaling of the dynamic
variables and energy)
\begin{equation}\label{2.1.6}
  \left( {x,y} \right) = l_{0i} \left( {\bar x,\bar y} \right),
  \left( {p_x ,p_y } \right) = p_{0i} \left( {\bar p_x ,\bar p_y } \right);
  E = \varepsilon _{0i} \bar E
\end{equation}
for these three cases is determined by the following parameters

\[ l_{01}  = \frac{a} {b},\quad p_{01}  = \sqrt {m\frac{{a^3 }}
{{b^2 }}} ,\quad \varepsilon _{01}  = \frac{{a^3 }} {{b^2 }}
\]

\[ l_{02}  = \sqrt {\frac{a} {c}} , \,\, p_{02}  = \sqrt {m
\frac{{a^2 }} {c}} ,\,\, \varepsilon _{02}  = \frac{{a^2 }}{c}
 \]

\begin{equation}\label{2.1.7}
l_{03}  = \frac{b} {c},\quad p_{03}  = \sqrt {m\frac{{b^4 }} {{c^3
}}} ,\quad \varepsilon _{03}  = \frac{{b^4 }} {{c^3 }}
\end{equation}

The reduced Hamiltonian for these three variants of scaling is the
following
\begin{equation}\label{2.1.8}
 \bar H_i \left( {\bar p_x ,\bar p_y ,\bar x,\bar y;W \equiv \frac{{b^2 }}
{{ac}}} \right) = \frac{{\bar p_x^2  + \bar p_y^2 }} {2} + \bar
U_i \left( {\bar x,\bar y,W} \right)
\end{equation}
where
\[  \bar {U}_1 \left( {\bar x,\bar y;W} \right) = \frac{1}
{2}\left( {\bar x^2  + \bar y^2 } \right) + \left( {\bar x^2 \bar
y - \frac{1} {3}\bar y^3 } \right) + \frac{1} {W}\left( {\bar x^2
+ \bar y^2 } \right)^2
\]
\[\bar U_2 \left( {\bar x,\bar y;W} \right) = \frac{1} {2}\left(
{\bar x^2  + \bar y^2 } \right) + \sqrt W \left( {\bar x^2 \bar y
- \frac{1} {3}\bar y^3 } \right) + \left( {\bar x^2  + \bar y^2 }
\right)^2\]
\begin{equation}\label{2.1.9}
\bar {U}_3 \left( {\bar x,\bar y,W} \right) = \frac{1}
{{2W}}\left( {\bar x^2  + \bar y^2 } \right) + \left( {\bar x^2
\bar y - \frac{1} {3}\bar y^3 } \right) + \left( {\bar x^2  + \bar
y^2 } \right)^2
\end{equation}
The transition $ i \to k $ between the different variants of
scaling is realized with the help of the substitution
\begin{equation}\label{2.1.10}
  \left( {\bar x,\bar y} \right) \to \left( {\bar x,\bar y} \right)\frac{{l_{ok} }}
{{l_{0i} }},\;\left( {\bar p_x ,\bar p_y } \right) \to \left(
{\bar p_x ,\bar p_y } \right)\frac{{p_{ok} }} {{p_{0i} }};\;\bar E
\to \frac{{\varepsilon _{0k} }} {{\varepsilon _{0i} }}\bar E
\end{equation}
where the parameters of transformation are defined by the
relations (\ref{2.1.7}).

At any variant of scaling the reduced Hamiltonian and
corresponding equations of motion depend only on parameter $W$.
This is the unique dimensionless parameter, which can be
constructed from the dimensional values $a,b,c.$ It means that for
each trajectory of initial "physical" Hamiltonian (\ref{2.1.3})
with energy $E$ , corresponds the unique trajectory of
one-parameter Hamiltonian (\ref{2.1.8}) with energy $\bar E =
E/\varepsilon _{0i} $ . While for each trajectory of the reduced
Hamiltonian $ \bar H_i (W)$ with  energy $ \bar E$ corresponds the
whole set of "physical" trajectories with energy $ E = \varepsilon
_{0i} \bar E $, which are generated by Hamiltonians $H(a,b,c)$
 with parameters satisfying the conditions $b^2 /ac = W$.

\subsection{ Geometry of potential energy surface}

Now, let's investigate the geometry of two-dimensional set of the
potential functions \\ $U(x,y;W)$ . The set of solutions of the
system of equations
\begin{equation}\label{2.2.1}
  U'_x  = 0,\quad U'_y  = 0,\quad \det \hat S = 0
\end{equation}
where $\hat S$ is the matrix of stability
\begin{equation}\label{2.2.2}
\hat S = \left| {\begin{array}{*{20}c}
   {U''_{xx} } & {U''_{xy} }  \\
   {U''_{xy} } & {U''_{yy} }  \\
 \end{array} } \right|
\end{equation}
serves by separatrix in the space of the parameters and divides it
into the regions, where the potential function is structurally
stable. The number and the nature of critical points change at the
transition through separatrix $W = 16$ and under the change of
$signW$ (i.e. under the change of $sign\;a;\;c$ is always
positive). The critical points of  PES for each structurally
stable  regions are given in Table 2.2.1.

\begin{center}
Table 2.2.1.  Number of critical points in different ranges.
\end{center}
\begin{center}
\begin{tabular}{|c|c|c|c|c|c|}  \hline
   & range & critical points & saddles & minima & maxima \\\hline
  I & $0<W<16$ & $1$ & $0$ & $1$ & $0$ \\\hline
  II & $W>16$ & $7$ & $3$ & $4$ & $0$ \\\hline
  III & $W<0$ & $7$ & $3$ & $3$ & $1$ \\ \hline
\end{tabular}
\end{center}

The corresponding lines of the level of the potential are
represented in Fig.1.
\begin{figure}
\centering
\includegraphics[height=4.5cm]{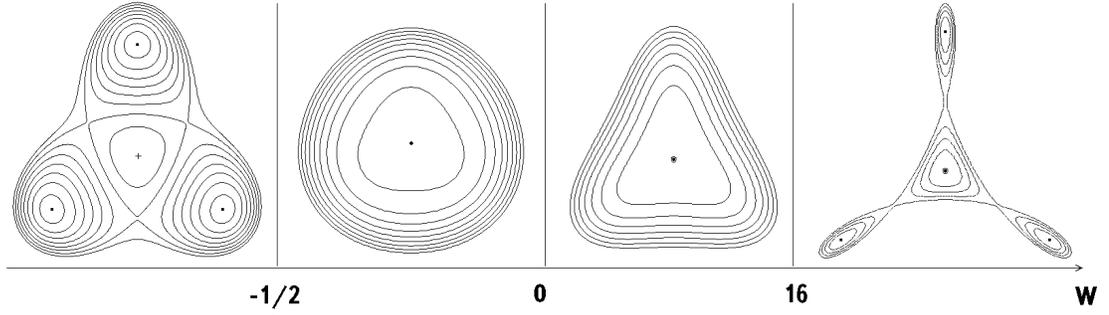}
\caption{The lines of the level of PES for different structurally
stable regions.} \label{fig.1}
\end{figure}

The coordinates of critical points are determined by
\begin{equation}\label{2.2.3}
\begin{gathered}
  \bar x_1  = 0,\quad y_1  = 0 \hfill \\
  \bar x_{2,3}  = 0,\quad y_{2,3}  = \frac{{\alpha _i }}
{8}\left( {1 \pm \sqrt {1 - \frac{{16}}
{W}} } \right) \hfill \\
  x_{4 - 7}  =  \pm \sqrt 3 y_{4 - 7} ,\quad y_{4 - 7}  =  - \frac{{\alpha _i }}
{{16}}\left( {1 \pm \sqrt {1 - \frac{{16}}
{W}} } \right) \hfill \\
\end{gathered}
\end{equation}
where $a_1  = W,a_2  = \sqrt W ,a_3  = 1.$

The region of the space of parameters $0 < W < 16$ includes the
potentials possessing the unique extremum - the minimum in the
origin of coordinates, corresponding to the spherically symmetric
equilibrium shape of nucleus. The region $W > 16$ includes the
potential surfaces with four minima. The central minimum, with $x
= y = 0$ , corresponds to the spherically symmetric equilibrium
shape of nucleus, but three peripheral minima correspond to
deformed shape. At last, in the region $W < 0$ $(a < 0)$ we run
into the potentials, describing the nuclei, which are deformed in
the ground state and even don't have the quasistable spherically
symmetric excited state. This region can be divided into two
subregions. In the space of parameters the boundary between these
subregions $W =  - 1/2$ represents the geometric locus, where the
both eigenvalues of matrix of stability turn into zero
simultaneously.
\begin{figure}
\centering
\includegraphics[height=9cm]{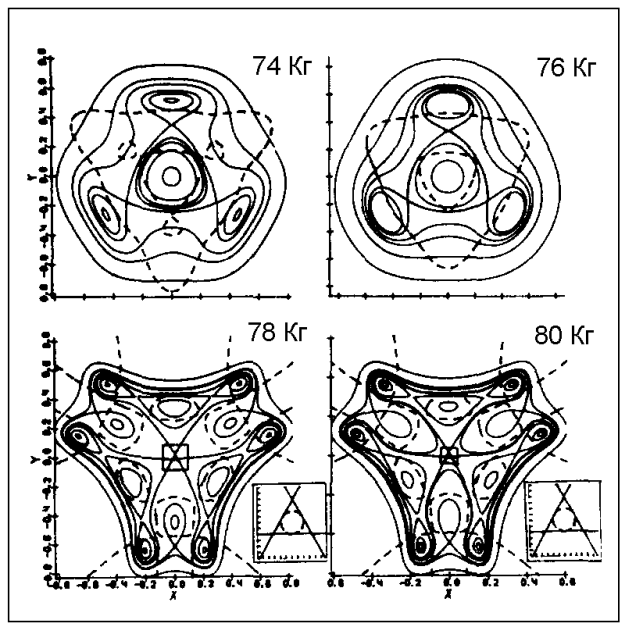}
\caption{The surfaces of potential energy of Krypton isotopes.}
\label{fig.2}
\end{figure}
Seiwert,Ramayya and Maruhn \cite{29} restored the parameters of
the Hamiltonian of the quadrupole oscillations, including the
sixth degree members in deformation for isotopes $Kr^{74,76,78,80}
$ . The big experimental values of energy of the first $2^ +  $ -
states for nuclei $Kr^{74,76} $ indicate spherical shape of
nucleus surface, while the probabilities of the electromagnetic
transitions $2^ + \to 0^ +  $ and very low energies of the first
rotational states (recalculating to one and the same number of
nucleons, this energy for isotopes $Kr^{74} $ is essentially lower
than the lowest from the known energy for nucleus $Pu^{240} $ -
$42.8\;kev$ ) imply the possibility of "superdeformation"
\cite{30}. The nonlinear effects, which are connected with the
geometry of PES, must be exhibited in the superdeformed nuclei
even at the relatively low energies of excitation. The potential
surfaces of Krypton isotopes, which are calculated in \cite{29},
are presented in Fig.2.
As it is seen, the inclusion into the
consideration of the highest members of expansion in deformation
leads to a considerable complication of the geometry of PES: for
all considered Krypton isotopes we run into the PES of complicated
topology with many local minima. It's clear, that in any degree in
deformation of the PES $C_{3\nu } $
 - symmetry is preserved.

\subsection{Critical energy of transition to chaos}

If we understand under the stochastization the process of
appearance in the system of statistical properties in  consequence
of local instability, we obtain a tempting possibility of
identifying the values of parameters, that lead to the local
instability in the system, with the boundary of transition to
chaos. Unfortunately, the criteria of the stochasticity of the
similar type, i.e. based on the investigation  of the local
instability, posses the innate lack: the lost of stability  of the
regular motion does not obligatory leads to chaos. Generally
speaking, instead of this the transition to another more
complicated type of regular motion is possible. Besides that, the
statement, that the local instabilities define the global dynamics
of the system is disputed. Separate details of the derivation of
the concrete criteria of the stochasticity provoke several
objections. In spite of these serious lacks the available
experience allows us to state, that the criteria of the similar
type in the aggregate with numerical experiment essentially
facilitate the analysis of the many-dimensional nonlinear motion.

A large variety of criteria of the stochasticity is based on the
direct evaluation of the rate of divergence of the initially close
trajectories $\left\{\vec q_1 (t),\vec p_1 (t)\right\}$ and
$\left\{\vec q_2 (t),\vec p_2 (t) \right\}$. Linearized equation
of motion for a divergence
\begin{equation}\label{2.3.1}
 \vec \xi \left( t \right) = \vec q_1 \left( t \right) - \vec q_2
 \left( t \right),\quad \vec \eta \left( t \right) =
 \vec p_1 \left( t \right) - \vec p_2 \left( t \right)
\end{equation}
assume the following form
\begin{equation}\label{2.3.2}
\dot{\vec\zeta}(t) = \vec \eta ,\quad \dot{\vec\eta}(t) = - \hat
S(t)\vec\zeta
\end{equation}
where $\hat S(t)$ is a matrix of stability, constructed from the
second derivatives of a potential $U(\vec q)$ and calculated along
the support trajectory $ \vec q_1 \left( t \right) $
\begin{equation}\label{2.3.3}
 S_{ij} \left( t \right) = \left. {\frac{{\partial ^2 U}}
{{\partial q_i \partial q_j }}} \right|_{\overrightarrow q  =
\overrightarrow {q_1 } \left( t \right)}
\end{equation}
The stability of motion of the dynamic system, described by the
Hamiltonian
\begin{equation}\label{2.3.4}
H\left( {\vec p,\vec q} \right) = \frac{{\vec p^2 }} {2} + U\left(
{\overrightarrow q } \right)
\end{equation}
is determined in $N$- dimensional case by $2N \times 2N$
 matrix
\begin{equation}\label{2.3.5}
\hat \Gamma  = \left| {\begin{array}{*{20}c}
   {\hat O} & {\hat I}  \\
   { - \hat S\left( t \right)} & {\hat O}  \\
 \end{array} } \right|
\end{equation}
where $\hat O$ and $\hat I$ is a zero and a unit$N \times N$
matrices. If even one of the eigenvalues $\lambda _i $ of the
matrix $\hat \Gamma $ is real, then the divergence of the
trajectory increases exponentially and the motion is unstable. The
imaginary eigenvalues correspond to the stable motion. The
eigenvalues, and consequently the character of motion change with
the time.

The problem of the investigation of the stable motion can be
essentially simplified \cite{31}, if we assume the possibility of
a replacement time-dependent point of phase space $\overrightarrow
{q_1 } (t)$ by time-independent coordinate $\vec q$ . It reduces
the equations for variations (\ref{2.3.2}) to a system of
autonomous linear differential equations
\begin{equation}\label{2.3.6}
\dot {\vec \xi  }= \vec \eta \quad ;\quad \dot {\vec \eta } =  -
\hat S\left( {\overrightarrow q } \right)\overrightarrow \xi
\end{equation}
An equation for Lyapunov exponents l which determine the character
of motion
\begin{equation}\label{2.3.7}
 \left| {\det \hat \Gamma  - \lambda \hat I} \right| = 0
\end{equation}
for the system with two degrees of freedom (the system of our
interest) has the following solution
\begin{equation}\label{2.3.8}
\lambda _{1,2,3,4}  =  \pm \left[ { - \beta  \pm \sqrt {\beta ^2 -
4\gamma } } \right]
\end{equation}
where
\begin{equation}\label{2.3.9}
\begin{gathered}
  \beta  = Sp'\hat S = U_{xx}  + U_{yy} , \hfill \\
  \gamma  = \det \hat S = U_{xx} U_{yy}  - U_{xy}^2  \hfill \\
\end{gathered}U_{xx}  \equiv \frac{{\partial ^2 U}}
{{\partial x^2 }},U_{yy}  \equiv \frac{{\partial ^2 U}} {{\partial
y^2 }},U_{xy}  \equiv \frac{{\partial ^2 U}} {{\partial x\partial
y}}.
\end{equation}
Here we will assume that $\beta  > 0$. Then, providing that
$\gamma  > 0$ , Lyapunov exponents  are purely imaginary and the
motion is stable. With $\gamma  < 0$ the pair of roots becomes
real and it leads to exponential divergence of close trajectories,
i.e. to the instability of motion.

Now let's remind several known facts from the theory of surfaces
\cite{32}. Gaussian curvature of a surface is equal to the ratio
of the determinants of the second $ \hat b$ and the first $ \hat
g$ quadratic forms
\begin{equation}\label{2.3.10}
 K = \frac{{b_{11} b_{22}  - b_{12}^2 }}
{{g_{11} g_{22}  - g_{12}^2 }} \Rightarrow  = \frac{{\det \hat b}}
{{\det \hat g}}
\end{equation}
In particular, if the surface is given in the form of the graph $z
= U\left( {x,y} \right)$ , then
\begin{equation}\label{2.3.11}
  \begin{gathered}
  \det \hat b = \frac{{U_{xx} U_{yy}  - U_{xy}^2 }}
{{1 + U_x^2  + U_y^2 }} \hfill \\
  \det \hat g = 1 + U_x^2  + U_y^2  \hfill \\
\end{gathered}
\end{equation}
and therefore
\begin{equation}\label{2.3.12}
K\left( {x,y} \right) = \frac{{U_{xx} U_{yy}  - U_{xy}^2 }}
{{\left( {1 + U_x^2  + U_y^2 } \right)^2 }}
\end{equation}
The Gaussian curvature $K$ can be represented in the form of the
product of the so-called principal curvatures $d_1 $ and $d_2 $ ,
which are the eigenvalues of the pair of the matrices $ \hat b $
and $ \hat g$ and are the solutions of the equation
\begin{equation}\label{2.3.13}
 \det \left( {\hat b - \delta \hat g} \right) = 0
\end{equation}
The sum of $d_1  + d_2 $ is called an average curvature of the
surface. We will enlarge on the geometric sense of the Gaussian
curvature. Let's choose the orthonormalized frame $(x,y,z)$ for
given point $x_0 ,y_0 $ of surface, where $z$ axis is a normal to
the surface. Then locally the surface will be written as $z =
U(x,y)$ and $U_x'  = U_y'  = 0$ in this point. Therefore
\begin{equation}\label{2.3.14}
\delta _1 \delta _2  = K = U_{xx} U_{yy}  - U_{xy}^2
\end{equation}
We will consider three possible cases

1) $ K > 0,\delta _1  > 0,\delta _2  > 0 $ (minimum of the
function $U(x,y)$ with $x = x_0 ,y = y_0 $ ,

2) $ K > 0,\delta _1 < 0,\delta _2  < 0$ (maximum of the function
$U(x,y)$ with $x = x_{_0 } ,y = y_0 $ ,

3) $ K < 0,\delta _1  < 0,\delta _2  > 0$ or vice versa (saddle
point of the function $U(x,y)$ with $x = x_0 ,y = y_0 $ .

At $K>0$ the surface is locally situated on one side of tangential
plane to the investigating point. At $K < 0$ the surface
necessarily crosses the tangential plane as close as possible to
the point of tangency. If the Gaussian curvature is positive
everywhere, then this surface is strictly convex.

Recently, in different sections of physics a definite interest has
grown to the surfaces which everywhere possess a negative
curvature \cite{33}. Such surfaces in the neighbourhood of any
point behave as in the neighbourhood of the hyperbolic singular
point. Now let's return to the expression of Liapunov exponents in
the case of two-dimensional potential surfaces. Comparing the
(\ref{2.3.9}) and (\ref{2.3.12}) expressions we notice that the g
sign coincides with the sign of Gaussian curvature of the PES.
This association suggests \cite{34,35} the possibility of the
existence of the following scenario of the transition from regular
to chaotic motion, based on the investigation of Gaussian
curvature of the PES.

At low energies the motion near the minimum of the potential
energy, where the curvature is obviously positive, is periodic or
quasiperiodic in character and is separated from the instability
region by the zero curvature line. As the energy grows, the
"particle" will stay for some time in the negative-curvature
region of the PES where initially close trajectories exponentially
diverge. At large time these results in the motion which imitate a
random one and is usually called stochastic. According to this
stochastization scenario, the critical energy of the transition to
chaos, $E_{cr} $ , coincides with the lowest energy on the zero
curvature line
\begin{equation}\label{2.3.15}
E_{cr}  = U_{\min } (K = 0)
\end{equation}
In the subsequent discussion we will reference to this statement,
as to the negative curvature criterion (NCC) .

Now we shall concentrate our attention on one of the first (1959
year) widely used criterion of transition to chaos, the so-called
overlap resonance's criterion (ORC) \cite{36} (the criterion of
Chirikov). The essence of this criterion is easier to explain by
the example of one-dimensional Hamiltonian system, which is
subjected to monochromatic periodic perturbation. This one is the
simplest Hamiltonian system which assumes the chaotic behaviour
\begin{equation}\label{2.3.16}
H = H_0 \left( {p,x} \right) + Fx\cos \Omega t
\end{equation}
For unperturbed system we can always introduce the variables
action-angle $\left( {I,\theta } \right)$ in which
\begin{equation}\label{2.3.17}
 H = H_0 \left( I \right) + \sum\limits_{k =  - \infty }^\infty
{x_k \left( I \right)\cos \left( {k\theta  - \Omega t} \right)}
\end{equation}
where
\begin{equation}\label{2.3.18}
x_k \left( I \right) = \frac{1} {{2\pi }}\int\limits_0^{2\pi }
{d\theta e^{ik\theta } x\left( {I,\theta } \right)}
\end{equation}

In new variables the scenario of stochasticity, on which the
overlap resonance's criterion (ORC) is based on, is the following.
An external field which is periodic in time induces a dense set of
resonances in the phase space of a nonlinear conservative
Hamiltonian system. The positions of these resonances, $I_k $ ,
are determined by the resonance condition between the
eigenfrequency $ \omega _k \left( I \right) = \frac{{\partial H_0
}} {{\partial I_k }}$ and frequency of the external perturbation,
$\Omega $ . For very weak external fields the principle resonance
zones remain isolated. As the amplitude $F$ of external field is
raised, the widths $W_k $ of the resonance zones increase
\begin{equation}\label{2.3.19}
 W_k  = \left. {4\left( {\frac{{Fx_k }}
{{\omega '\left( I \right)}}} \right)^{{\raise0.7ex\hbox{$1$}
\!\mathord{\left/
 {\vphantom {1 2}}\right.\kern-\nulldelimiterspace}
\!\lower0.7ex\hbox{$2$}}} } \right|_{I = I_k }
\end{equation}
and at $F > F_{cr} $ resonances overlap. When this overlap occurs,
i.e. under the condition
\begin{equation}\label{2.3.20}
 \frac{1}
{2}\left( {W_k  + W_{k + 1} } \right) = \left| {I_k  - I_{k + 1} }
\right|
\end{equation}
it is said that there is transition to a global stochastic
behaviour in the corresponding region of the phase space. In other
words the ORC postulates, that the last invariant KAM surface,
which separates the neighbouring resonances, is destroyed in the
moment of contact of unperturbed separatrices of these resonances.
In other words , the averaged motion of the system in the
neighbourhood of the nonlinear isolated resonance on the plane of
the variables action-angle is similar to the particle behavior in
the potential well. Several isolated resonances correspond to
several isolated potential wells. The overlap of the resonances
means, that there is such an approach of the potential wells,
wherein the random walk of a particle between these wells is
possible.

The outlined scenario can easily be "corrected" for the
description of the transition to chaos in the conservative system
with several degrees of freedom. The condition of the resonance
between the eigenfrequency and the frequency of external field
must be replaced by the condition of the resonance between the
frequencies, which correspond to different degrees of freedom
\begin{equation}\label{2.3.21}
 \sum {m_i \frac{{\partial H_0 }}
{{\partial I_i }} = 0}
\end{equation}
The role of the amplitude of the external field in this case plays
the intensity of the interaction between different degrees of
freedom , i.e. the measure of nonlinearity of the original
Hamiltonian. This parameter is usually the energy of system.

This method must be slightly modified for the systems with the
unique resonance. In this case the origin of the large-scale
stochasticity is connected \cite{37} with the destruction of the
stochastic layer near the separatrix of this unique resonance. The
essence of the modification consists in the approximate reduction
of the original Hamiltonian in the neighbourhood of resonance to
the Hamiltonian of nonlinear pendulum, which interacts with
periodic perturbation
\begin{equation}\label{2.3.22}
 H\left( {v,x,\tau } \right) = \frac{1}
{2}v^2  - M\cos x - P\cos k\left( {x - \tau } \right)
\end{equation}
The width w of the stochastic layer of the resonance is equal
to\cite{38}
\begin{equation}\label{2.3.23}
w \approx {\raise0.7ex\hbox{${pe^{ - {\raise0.7ex\hbox{$1$}
\!\mathord{\left/
 {\vphantom {1 \rho }}\right.\kern-\nulldelimiterspace}
\!\lower0.7ex\hbox{$\rho $}}} }$} \!\mathord{\left/
 {\vphantom {{pe^{ - {\raise0.7ex\hbox{$1$} \!\mathord{\left/
 {\vphantom {1 \rho }}\right.\kern-\nulldelimiterspace}
\!\lower0.7ex\hbox{$\rho $}}} } {M\rho ^{2k + 1}
}}}\right.\kern-\nulldelimiterspace} \!\lower0.7ex\hbox{${M\rho
^{2k + 1} }$}}
\end{equation}
where
\begin{equation}\label{2.3.24}
\rho  = {\raise0.7ex\hbox{${2M^{{\raise0.7ex\hbox{$1$}
\!\mathord{\left/
 {\vphantom {1 2}}\right.\kern-\nulldelimiterspace}
\!\lower0.7ex\hbox{$2$}}} }$} \!\mathord{\left/
 {\vphantom {{2M^{{\raise0.7ex\hbox{$1$} \!\mathord{\left/
 {\vphantom {1 2}}\right.\kern-\nulldelimiterspace}
\!\lower0.7ex\hbox{$2$}}} } {\pi
k}}}\right.\kern-\nulldelimiterspace} \!\lower0.7ex\hbox{${\pi
k}$}}
\end{equation}
If $p/M$ has the power $\rho ^S $ , then
\begin{equation}\label{2.3.25}
w \approx \rho ^{ - \lambda } e^{ - {\raise0.7ex\hbox{$1$}
\!\mathord{\left/
 {\vphantom {1 \rho }}\right.\kern-\nulldelimiterspace}
\!\lower0.7ex\hbox{$\rho $}}}
\end{equation}
where
\begin{equation}\label{2.3.26}
\lambda  = 2k + 1 - s.
\end{equation}
At
\begin{equation}\label{2.3.27}
  \rho _i  = \frac{1} {\lambda }\left[ {1 - \left( {1 + \lambda }
\right)^{ - {\raise0.7ex\hbox{$1$} \!\mathord{\left/
 {\vphantom {1 2}}\right.\kern-\nulldelimiterspace}
\!\lower0.7ex\hbox{$2$}}} } \right]
\end{equation}
the function $w(r)$ has a point of inflection. The fast growth of
$w$ allows us to determine the thresholds of the destruction of
the stochastic layer as a value
\begin{equation}\label{2.3.28}
  \rho _s  = \lambda ^{ - 2} \left[ {\left( {1 + \lambda } \right)^{{\raise0.7ex\hbox{$1$} \!\mathord{\left/
 {\vphantom {1 2}}\right.\kern-\nulldelimiterspace}
\!\lower0.7ex\hbox{$2$}}}  - 1} \right]^2
\end{equation}
This value $\rho $ is such that the tangent to the point $\rho _i
$ of the function $w$ crosses the axis $\rho $ .

All three described criteria of stochasticity will be used below
to determine the critical parameters of the transition to chaos.

\subsection{ Numerical results versus analytical estimations}

Now let us turn to the analysis of the solutions of the equations
of motion, which are generated by the Hamiltonian (\ref{2.1.3}).
As mentioned above in the section 2.2, the geometry of the PES for
the regions

I    $0 < W < 16$

II        $W > 16$

III  $W < 0(a < 0)$

is essentially different. Undoubtedly the specific character of
the PES must be manifested in behaviour of the solutions of the
equations of motion. Therefore we shall analyze each of the
mentioned regions separately. Notice, that parameters of the
Hamiltonian $a,b,c$ were estimated in different phenomenological
models \cite{39}. They change in such wide limits that for real
nuclei parameter $ W = {\raise0.7ex\hbox{${b^2 }$}
\!\mathord{\left/
 {\vphantom {{b^2 } {ac}}}\right.\kern-\nulldelimiterspace}
\!\lower0.7ex\hbox{${ac}$}}$ can belong to each of the regions
I,II,III.

\subsubsection{ Region $0 < W < 16$ : potentials with unique
extremum}

It is the simplest region for the analysis in which good agreement
between the boundary of the transition to chaos, observed in
numerical experiments and analytical evaluations, can be obtained
with the help of the simplest of the criterion of the transition
to chaos - the negative curvature criterion. Once more we remind,
that according to this criterion the mechanism of the generation
of the local instability consists in hit of the particle in the
region of negative Gaussian curvature of the PES and critical
energy of the transition to chaos coincides with the minimum value
of the energy on the zero Gaussian curvature line. The equation of
the last one in re-scale coordinates (further, we shall use the
third variant of scaling (\ref{2.1.9})) is
\begin{equation}\label{2.4.1}
  \frac{1}
{{48W^2 }} - \frac{{W - 4}} {{12W}}\left( {\bar x^2  + \bar y^2 }
\right) + \left( {\bar x^2  + \bar y^2 } \right)^2  - \left( {\bar
x^2 \bar y - \frac{1} {3}\bar y^3 } \right) = 0
\end{equation}
and potential energy on the zero curvature line is
\begin{equation}\label{2.4.2}
U^{K = 0} \left( {\bar x,\bar y} \right) = \frac{1} {{48W^2 }} -
\frac{{W - 10}} {{12W}}\left( {\bar x^2  + \bar y^2 } \right) +
2\left( {\bar x^2  + \bar y^2 } \right)^2
\end{equation}
As it is seen, the zero curvature line conserves the symmetry of
the PES (see Fig.3.). Therefore, the minimum of the energy on the
zero curvature line must lie either on the straight line $x = 0$
or on the straight lines, obtained from it with the help of the
transformations of the symmetry of the discrete group $C_{3\nu }
$.
\begin{figure}
\centering
\includegraphics[height=8cm]{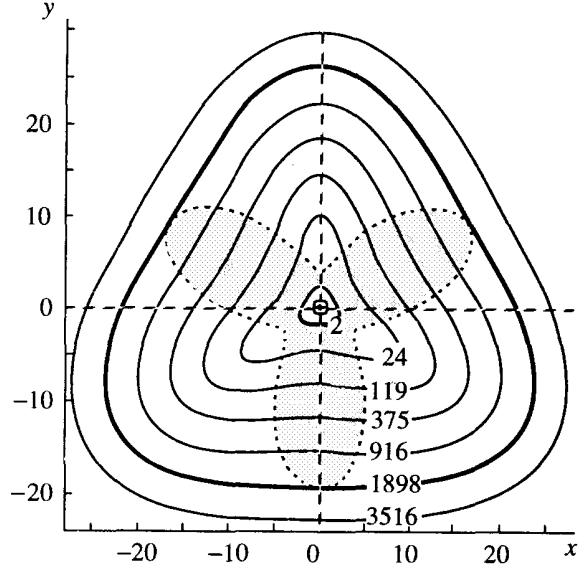}
\caption{Isolines of the PES for $W = 13$ . Zero curvature line
$K\left( {x,y} \right) = 0$ (dashed line) is shown on the
background of the level lines, the range $K\left( {x,y} \right) <
0$ is shaded.} \label{fig.3}
\end{figure}

At $0 < W < 4$ Gaussian curvature of the PES is positive
everywhere. At $4 < W < 12$ the subregion of the negative
curvature is localized on the straight line (in the plane) $x = 0$
in the interval
\begin{equation}\label{2.4.3}
   - \frac{1}
{4}\left( {1 + \sqrt {1 - \frac{4} {W}} } \right) < \bar y <  -
\frac{1} {4}\left( {1 - \sqrt {1 - \frac{4} {W}} } \right)
\end{equation}
and in the subregion $12 < W < 16$ the negative curvature appears
at $ \bar y > 0$ in the interval
\begin{equation}\label{2.4.4}
\frac{1} {{12}}\left( {1 - \sqrt {1 - \frac{{12}} {W}} } \right) <
\bar y < \frac{1} {{12}}\left( {1 + \sqrt {1 - \frac{{12}} {W}} }
\right)
\end{equation}
In Fig. 4. the profiles $ U\left( {\bar x = 0,\bar y} \right)$ of
the PES are shown for the three considered subregions, and the
intervals of the negative curvature are shaded.

Thus, according to the considered scenario of the stochastization
in the subregion $0 < W < 4$ the motion must remain regular for
all energies. According to the (\ref{2.3.15}) in the neighbourhood
of the energy
\begin{equation}\label{2.4.5}
E_{cr}  = U_{\min } \left( {K = 0} \right) = U\left( {x =
0,y_{cr1} } \right),y_{cr1}  =  - \frac{1} {4}\left( {1 - \sqrt {1
- \frac{4} {W}} } \right)
\end{equation}
the transition to the global stochasticity should be expected in
the subregion $4 < W < 12$. Finally, in the sub region $12 < W <
16$ this transition must be observed in the neighbourhood of the
energy
\begin{equation}\label{2.4.6}
 E_{cr}  = U_{\min } \left( {K = 0} \right) = U\left( {x = 0,y_{cr2} } \right),y_{cr2}  = \frac{1}
{{12}}\left( {1 - \sqrt {1 - \frac{{12}} {W}} } \right)
\end{equation}
In this case we have used that for all $12 < W < 16$ $U(x =
0,y_{cr1} ) > U(x = 0,y_{cr2} )$ .

\begin{figure}[t]
\centering
\includegraphics[height=10cm]{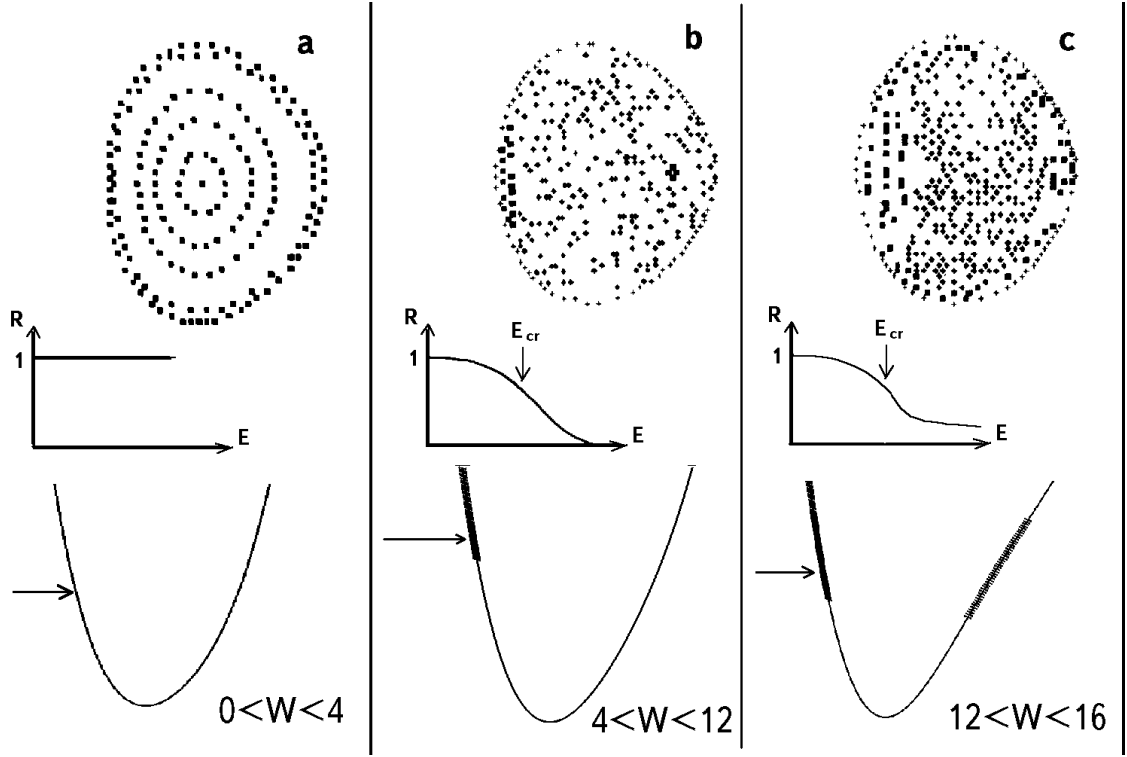}
\caption{The profiles $ U\left( {\bar x = 0,\bar y} \right)$ of
the PES for the sub regions $0 < W < 4,\;4 < W < 12$ and $12 < W <
16$ . The regions of the negative curvature are shaded. In the top
part of the Figure - the Poincare surfaces of section for the
energy values indicated by arrows, are represented.} \label{fig.4}
\end{figure}
These predictions must be compared with the numerical solutions of
the equations of motion generated by the Hamiltonian
(\ref{2.1.8}).

Perhaps, the simplest numerical method of the detection of
stochasticity is the analysis of the Poincare surfaces of section.
As it is well known [40] the Hamiltonian system with $N$ degrees
of freedom is integrable only in the case of the existence of $N$
independent univalent integrals of motion. If the number of the
integrals of motion is less than the number of degrees of freedom,
then the dynamic chaos is possible in the system. In the case of
the existence of one or more additional (except energy) integrals
of motion, the points of intersection of the phase trajectory with
the arbitrary chosen plane (it is just Poincare surface of
section) lie on the variety, the dimension of which is smaller
than $2N - 2$ . Otherwise they will fill all the $2N - 2$
dimensional isoenergetic volume (all $2N - 2$ isoenergetic
surface). Therefore the analysis of the Poincare surfaces of
section allows us to establish the fact of the existence of the
additional integrals of motion and, consequently, to elucidate
which type of motion is realized in the system with the given
initial condition. The analysis of Poincare surfaces of section is
especially effective for the system with two degrees of freedom,
the phase space of which is four-dimensional. In view of  the
conservation of energy, the trajectory of the particle lies on
three-dimensional surface $ H\left( {p_x ,p_y ,x,y} \right) =const
$ . Excepted one of the variables, for example $p_x $ , we shall
consider the points of intersection of phase trajectory with the
plane $x = 0$ . In common case they will be chaotic distributed on
some part of the plane$(p_y ,y)$ restricted by separatrix. In the
case of existence of the additional integral of motion $ I\left(
{p_x ,p_y ,x,y} \right) = const$ the totality of the consecutive
intersections with the chosen plane lies on some curve $f(p_y ,y)
= const$ . By contrast, chaotic trajectories are identified by the
fact that they show no orderly pattern on the Poincare surface of
section. As we move from a regular  regime into the chaotic one,
disorderly trajectories seem to appear first near separatrices
between different types of regular orbits. They become visible
only in small regions, but as we move further into the chaotic
regime, they cover most or almost all of the surface of section
until eventually no regular  trajectories are visible.

The analysis of Poincare surfaces of section, obtained by the
numerical integration of Hamiltonian equations of motion for the
PES with unique extremum $(0 < W < 16)$ , leads to the following
results:

1) At low energies $(E <  < E_{cr} )$ in the neighbourhood of
minimum for all considered values $W$ , the motion remains
regular. The regularity of the motion with low energies  is a
straight consequence of the KAM theorem \cite{41,42}, which
states, that the majority of the regular trajectories of the
unperturbed systems remain regular under sufficiently small
perturbation. It is clear that this is accorded with the
positiveness of Gaussian curvature in the neighbourhood of any
minimum.

2) In the interval $0 < W < 4$ for all energies the motion remains
regular (see Fig. 4a.). This can be explained by the positiveness
of the Gaussian curvature of the PES in this region of the
parameter $W$ .

3) In the interval $4 < W < 16$ as the energy increases, the
gradual transition from the regular (quasi periodic) motion to
chaotic one (see Fig. 4b.) is observed. Moreover in the subregion
$4 < W < 12$ the critical energy, observing in the Poincare
surface of section, is close to (\ref{2.4.5}); as in the subregion
$12 < W < 16$ the transition to the global stochasticity is
observed in the neighbourhood of the energy, calculated according
(\ref{2.4.6}). This effect at $12 < W < 16$ is connected with the
beginning of the region of negative curvature at $y > 0$ located
below in the energy (see Fig. 4c.).

It is necessary to make one important remark. Analysis of the
Poincare surfaces of section allows to introduce the critical
energy of the transition to chaos, having determined it as the
energy in which the part of phase space with chaotic motion
exceeds certain arbitrary chosen value. Similar indetermination is
connected with the absence of the sharp transition to chaos for
any critical value of the perturbation to which integrated system
is undertaken. Therefore a certain caution is required in
comparing the "approximate" critical energy, obtained by any
variant of numerical simulation, with the "exact" value obtained
with the help of analytical estimations, i.e. on the base of
different criteria of stochasticity.

Based on this remark we can say, that in the case of one-well
potentials, the negative curvature criterion allows to make
reliable predictions about the possibility of the existence of
chaotic regimes in the considered region of the parameters; and
also to evaluate the region of energies at which the transition
regularity-chaos is performed.

\subsubsection{ Region $W > 16$ : the potentials with a few local
minima}

Now we are proceeding to the analysis of numerical solutions of
the equations of motion in region $W > 16$ . The geometry of the
PES, which is more complicated in comparison with the potentials
which have the unique extremum (see Table 2.2.1), assumes the
existence of several energies of the transition to chaos even for
the fixed set of parameters of the potential. It means, that for
such potentials the so-called mixed states \cite{43} must be
observed: at one and the same energy in different minima the
various dynamic regimes are realized. The Poincare surfaces of
section at different energies for Hamiltonian (\ref{2.1.8}) with
$W = 18$ are presented in Fig.5. This value $W$ provides equality
of depths for the central and peripheral minima. Taking into
account $C_{3\nu } $ symmetry of the PES, only one peripheral
minimum is presented in this Figure.
\begin{figure}
\centering
\includegraphics[height=8.5cm]{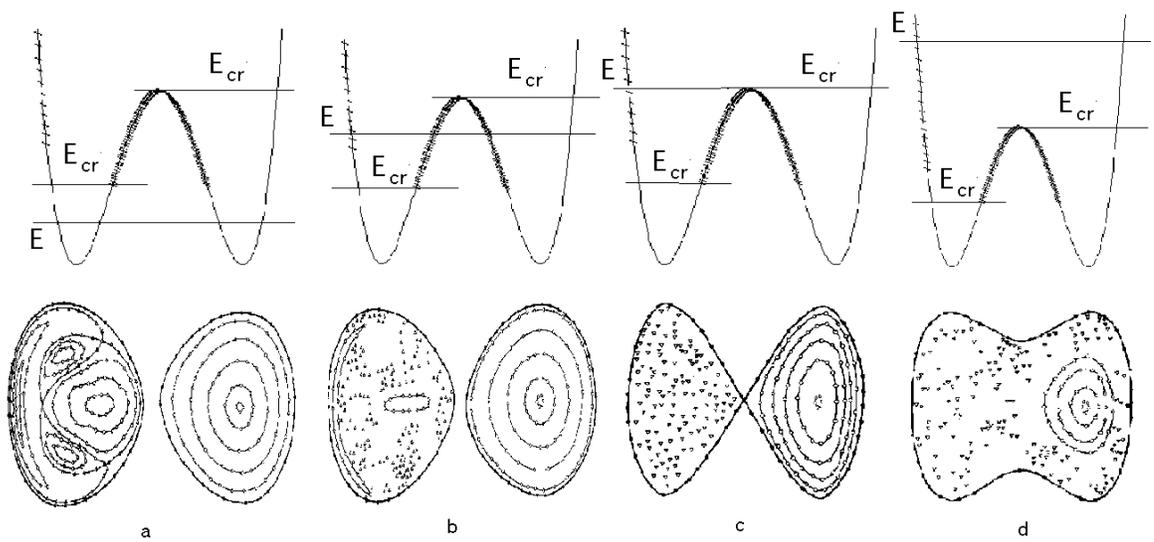}
\caption{The Poincare surfaces of section for the motion in the
potential with $W = 18$ at different energies: at the energy less
then critical, $E \ll E_{cr} $ , (b) at the energy above critical,
$E > E_{cr} $ , (c) at the energy equal to saddle, $E = E_S $ .and
(d) at the energy above the saddle, $E > E_S $ .. The profile
$U\left( {x = 0,y} \right)$ of potential is presented below. The
range of $K < 0$ is shaded.} \label{fig.5}
\end{figure}
The motion represented in Fig. 5.a  has clearly defined
quasiperiodic character both for the central (left minimum) and
for the peripheral (right) minima. Special attention must be given
to the distinction in structure of the Poincare surfaces of
section for different minima: the complicated structure with
several fixed points in left minimum and simple structure with the
unique fixed elliptic point in right one. The gradual transition
to chaos is observed with the increasing of energy, however the
change of the character of motion of the trajectories, localized
in certain minimum, is essentially different. Whereas there is the
gradual transition to chaos for the left well even for the energy
approximately equal to one-half of the saddle energy (Fig. 5.b),
and for the energy equal to the saddle one (Fig. 5c), practically
all initial conditions correspond to the chaotic trajectories, the
motion remains quasi periodic in the (second) right minimum at the
same energies. In this minimum the transition to global
stochasticity takes place only at the neighbourhood of the saddle
energy. In the right well the significant part of phase space
corresponding to quasi periodic motion conserves even at the
energy essentially exceeding the saddle one (Fig.5d). Figure 6.
shows a comparison of the critical energies of the transition to
chaos obtained according to the negative curvature criterion and
with the help of the analysis of the Poincare surfaces of section.
On the base of this comparison we can do the following
conclusions.
\begin{figure}
\centering
\includegraphics[height=8cm]{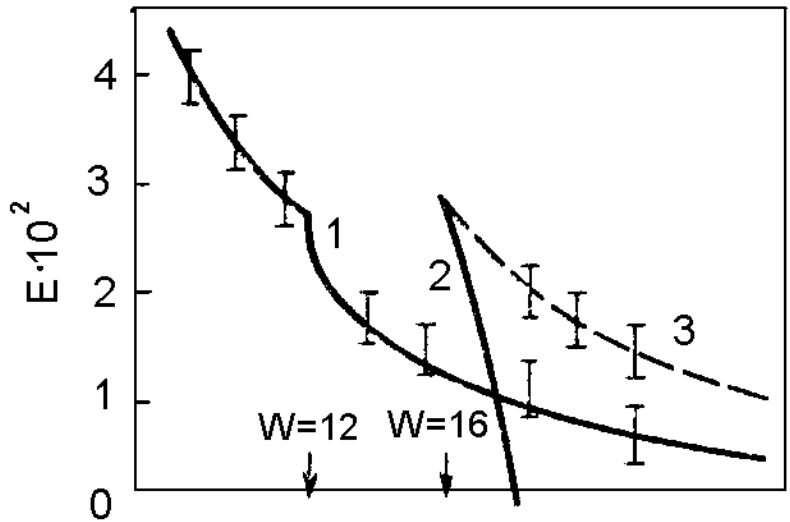}
\caption{The critical energy of the transition to chaos for
region: the negative curvature criterion (rigid line) and with the
help of the analysis of the Poincare section. } \label{fig.6}
\end{figure}

1) The mixed state is observed for considered potential for region
of energies $1/2E_S  < E < E_S (E_S  - $ saddle energy).

2) The critical energy, defined according to the negative
curvature criterion for the left well of potential, is in a good
agreement with obtained one by means of numerical integration of
the equations of motion and contradicts to the situation for the
right well, where the numerical simulation detects large-scale
chaos only at attainment of the saddle energy.

The mixed state, which is shown above for the potential of
quadrupole oscillations, is the representative state for the wide
class of two-dimensional potentials with a few local minima. We
shall show this by the example of the polynomial potentials of the
degree not higher than six, which are symmetric relatively to the
plane $x = 0$ . But even with such restrictions, the possible set
of the potential forms, depending (generally speaking) on 12
parameters, is too big. Remaining generality, we shall use the
methods of the theory of catastrophe \cite{44} in order to reduce
amount of calculations. According to the last one, a rather wide
class of polynomial potentials with the several local minima is
covered by the germs of the lowest umbilical catastrophes, of type
$D_5 ,D_4^ -  ,D_7 $ , subjected to the definite perturbations.
Let us notice, that the potential of type Henon-Heiles coincides
with the elliptic ombilic $D_4^ -  $ \cite{44} with precision to
linear terms of perturbation.

The mixed state is observed for all considered potentials of
umbilical catastrophes in the interval of energies $E_{_{cr} }  <
E < E_S $ (here $E_{cr} $ is the critical energy of the transition
to chaos determined by negative curvature criterion). The
transition to chaos (contrary to the negative curvature criterion)
is observed only at the reach of the saddle energy for the minima
possessing unique fixed elliptic point in the Poincare surface of
section, as in the case of the potential of quadrupole
oscillations. This contradiction makes one to refer to the
criteria of stochasticity described in the section 2.3., and based
on the theory of non-linear resonance.

Let us consider, for example, the Hamiltonian with the PES which
represents the germ of the catastrophe $D_5 $ with quadratic
perturbation
\begin{equation}\label{2.4.7}
  H = \frac{1}
{2}\left( {p_x^2  + p_y^2 } \right) + \left[ {\frac{1} {4}y^4  +
x^2 y + ax^2  - y^2 } \right]
\end{equation}
The geometry of the two-dimensional one-parameter potential
$U(x,y;a)$ at $a > \sqrt 2 $ is determined by five critical points
: by two minima of equal depth (which thereafter will be named
left and right wells) and by three saddles. The energy of the
saddle, situated at the origin of the coordinates and separating
the wells, does not depend on a and is equal to zero. Therefore,
the classical motion will be localized at the separate well at the
negative energies. The energies in all the saddles coincide with
each other at $a = 2$ (this case will be investigated in detail).
Now let us estimate the critical energy of the transition to chaos
by the negative curvature criterion. In this case the problem is
reduced to the search of the conditional extremum - minimum of the
potential energy on the zero curvature line. The last one is
described by equation
\begin{equation}\label{2.4.8}
  \left( {y + a} \right)\left( {3y^2  - 2} \right) - 2x^2  = 0
\end{equation}
At $a = 2$ we come to the value of the critical energy, which is
equal for the both wells
\[
E_{cr}  = U_{\min } \left( {K = 0} \right) =  -
{\raise0.7ex\hbox{$5$} \!\mathord{\left/
 {\vphantom {5 9}}\right.\kern-\nulldelimiterspace}
\!\lower0.7ex\hbox{$9$}}.
\]

The results of numerical integration of the equation of motion
generated by the Hamiltonian (\ref{2.4.7}) and presented in Fig.7
qualitatively coincide with the results obtained for the
Hamiltonian of quadrupole oscillations (Fig.5). Well manifested
mixed state is observed at the energies $ E \geqslant
{\raise0.7ex\hbox{$1$} \!\mathord{\left/
 {\vphantom {1 2}}\right.\kern-\nulldelimiterspace}
\!\lower0.7ex\hbox{$2$}}E_S $ in Poincare surfaces of section.

For the explanation of this phenomenon we shall consider the
dynamics in the variables angle-action. The Hamiltonian
(\ref{2.4.7}) in the system of coordinates with the origin in the
left (top sign) and in the right (bottom sign) has the following
form
\begin{equation}\label{2.4.9}
H = \frac{1} {2}\left( {\dot x^2  + \omega _1^2 x^2 } \right) +
\frac{1} {2}\left( {\dot y^2  + \omega _2^2 y^2 } \right) + x^2 y
\mp \sqrt 2 y^3  + \frac{1} {4}y^4
\end{equation}
where       $ \omega _1  = \left[ {2\left( {a \mp \sqrt 2 }
\right)} \right]^{{\raise0.7ex\hbox{$1$} \!\mathord{\left/
 {\vphantom {1 2}}\right.\kern-\nulldelimiterspace}
\!\lower0.7ex\hbox{$2$}}} ,\omega _2  = 2.$

Now we perform the canonical transformation to the variables
action-angle of the oscillator part of the Hamiltonian
\begin{equation}\label{2.4.10}
  \begin{array}{*{20}c}
   {x = \left( {\frac{{2I_1 }}
{{\omega _1 }}} \right)^{{\raise0.7ex\hbox{$1$} \!\mathord{\left/
 {\vphantom {1 2}}\right.\kern-\nulldelimiterspace}
\!\lower0.7ex\hbox{$2$}}} \cos \varphi _1 ,} & {y = \left(
{\frac{{2I_2 }} {{\omega _2 }}} \right)^{{\raise0.7ex\hbox{$1$}
\!\mathord{\left/
 {\vphantom {1 2}}\right.\kern-\nulldelimiterspace}
\!\lower0.7ex\hbox{$2$}}} \cos \varphi _2 ,}  \\
   {\dot x = \left( {2I_1 \omega _1 } \right)^{{\raise0.7ex\hbox{$1$} \!\mathord{\left/
 {\vphantom {1 2}}\right.\kern-\nulldelimiterspace}
\!\lower0.7ex\hbox{$2$}}} \cos \varphi _1 ,} & {\dot y = \left(
{2I_2 \omega _2 } \right)^{{\raise0.7ex\hbox{$1$}
\!\mathord{\left/
 {\vphantom {1 2}}\right.\kern-\nulldelimiterspace}
\!\lower0.7ex\hbox{$2$}}} \cos \varphi _2 .}  \\
 \end{array}
\end{equation}
\begin{figure}[t]
\centering
\includegraphics[height=7.5cm]{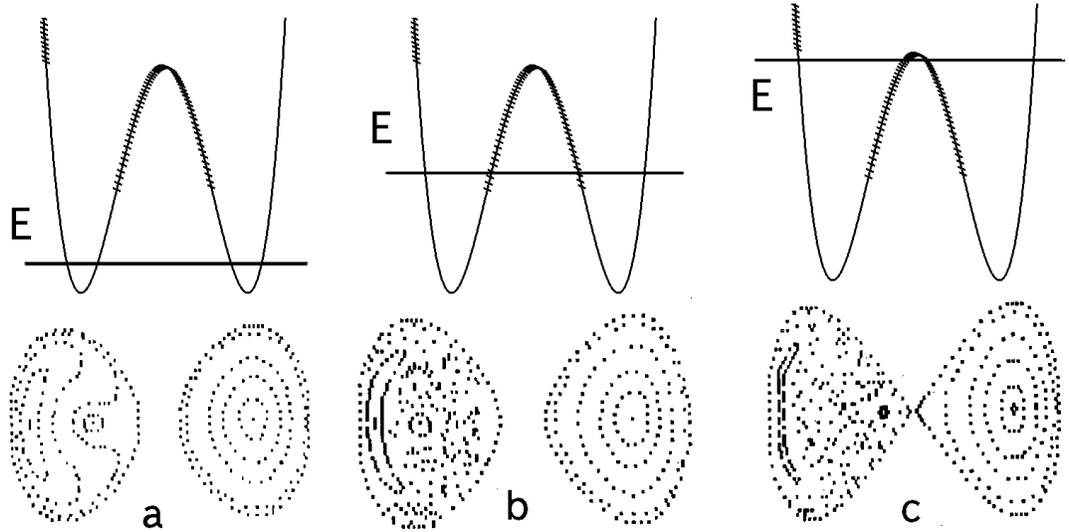}
\caption{ The results of numerical integration of the equation of
motion generated by the Hamiltonian with PES which represents the
germ of the catastrophe $D_5 $ with quadratic perturbation
(\ref{2.4.7}). In the top part - the profile $U\left( {x = 0,y}
\right)$ of potential. Below - the Poincare surfaces: (a) at the
energy less then critical, $E \ll E_{cr} $ , (b) at the energy
sligthly above critical, $E > E_{cr} $ and (c) at the energy equal
to saddle, $E = E_S $ . } \label{fig.7}
\end{figure}
In these variables the Hamiltonian (\ref{2.4.9}) has the following
form
\begin{equation}\label{2.4.11}
 \begin{gathered}
  H\left( {I_1 ,I_2 ,\varphi _1 ,\varphi _2 } \right) =
  H_0 \left( {I_1 ,I_2 } \right) + \sum\limits_{m_1 m_2
   \in y} {f_{m_1 m_2 } \left( {I_1 ,I_2 } \right)} \cos
   \left( {m_1 \varphi _1  + m_2 \varphi _2 } \right) \hfill \\
  y:\left[ {0,1} \right],\left[ {0,2} \right],\left[ {0,3}
   \right],\left[ {0,4} \right],\left[ {2,1} \right],\left[ {2, - 1}
    \right] \hfill \\
\end{gathered}
\end{equation}
where
\[H_0 \left( {I_1 ,I_2 } \right) = I_1 \omega _1  + I_2 \omega _2  +
\frac{3} {8}\frac{{I_2^2 }} {{\omega _2^2 }} - 1\]
\[f_{01}  = \frac{{I_1 }} {{\omega _1 }}\left( {\frac{{2I_2 }}
{{\omega _2 }}} \right)^{{\raise0.7ex\hbox{$1$} \!\mathord{\left/
 {\vphantom {1 2}}\right.\kern-\nulldelimiterspace}
\!\lower0.7ex\hbox{$2$}}}  \pm 3\left( {\frac{{I_2 }} {{\omega _2
}}} \right)^{{\raise0.7ex\hbox{$3$} \!\mathord{\left/
 {\vphantom {3 2}}\right.\kern-\nulldelimiterspace}
\!\lower0.7ex\hbox{$2$}}}\]
\[f_{02}  = \frac{1} {2}\frac{{I_2^2 }} {{\omega _2^2 }},f_{03}  =
\pm \left( {\frac{{I_2 }} {{\omega _2 }}}
\right)^{{\raise0.7ex\hbox{$3$} \!\mathord{\left/
 {\vphantom {3 2}}\right.\kern-\nulldelimiterspace}
\!\lower0.7ex\hbox{$2$}}}\]
\begin{equation}\label{2.4.12}
f_{04}  = \frac{1} {8}\frac{{I_2^2 }} {{\omega _2^2 }},f_{21}  =
f_{2, - 1}  = \frac{1} {2}\frac{{I_1 }} {{\omega _1 }}\left(
{\frac{{2I_2 }} {{\omega _2 }}} \right)^{{\raise0.7ex\hbox{$1$}
\!\mathord{\left/
 {\vphantom {1 2}}\right.\kern-\nulldelimiterspace}
\!\lower0.7ex\hbox{$2$}}}
\end{equation}
The term with indexes $r = (r_1 ,r_2 )$ is called the resonance
term for the given value of energy $E$ , if there are variables
action $ \left( {I_1^r ,I_2^r } \right)$ such that $ E = H_0
\left( {I_1^r ,I_2^r } \right) $ :
\begin{equation}\label{2.4.13}
r_1 \bar \omega _1 \left( {I_1^r ,I_2^r } \right) + r_2 \bar
\omega _2 \left( {I_1^r ,I_2^r } \right) = 0
\end{equation}
where
\begin{equation}\label{2.4.14}
\bar \omega _i  = {\raise0.7ex\hbox{${\partial H_0 }$}
\!\mathord{\left/
 {\vphantom {{\partial H_0 } {\partial I_i }}}\right.\kern-\nulldelimiterspace}
\!\lower0.7ex\hbox{${\partial I_i }$}},i = 1,2.
\end{equation}

It is possible to avoid problems of "small dominators" to perform
canonical transformation to the new action-angle variables which
eliminate the dependence of the angle in the lowest order in small
parameter identified with the energy at the energies corresponding
to the finite motion $( - 1 < E < 0)$ , if we are situated
sufficiently far from the resonance, i.e. for all $m_1 ,m_2 ,$
\begin{equation}\label{2.4.15}
  \left| {m_1 \bar \omega _1  + m_2 \bar \omega _2 } \right| \leqslant f_{m_1 m_2 }
\end{equation}

The result of this procedure is the overdetermination of an
integrable part $H_0 (I_1 ,I_2 )$ of the initial Hamiltonian and
the extension of the set J members depending on angles. In this
situation it is possible to meet one of the following three cases:

1) the resonance terms are, as before, absent in the region of
energies of our interest;

2) the unique resonance term appears;

3) several resonance terms appear.

In the first case we perform a new canonical transformation and
continue the procedure until we meet the situation 1) or 2). In
the second case the critical energy, at which the transition to
large-scale stochasticity takes place, can be determined by the
destruction stochastic layer method \cite{37}, while the overlap
resonances criterion can be used \cite{36} for the determination
of the critical energy in the third case.

Small detuning near the resonance
\begin{equation}\label{2.4.16}
  \left| {m_1 \bar \omega _1  + m_2 \bar \omega _2 } \right| \leqslant f_{m_1 m_2 }
\end{equation}
can be compensated by terms of the highest order obtained with the
help of repeated canonical transformation of non-resonance terms.
In the case when $a = 2$ for the resonance (2,-1) the condition
(\ref{2.4.16}) of small detuning is performed and procedure,
mentioned above, leads to
\begin{equation}\label{2.4.17}
 \begin{gathered}
  H_0  = I_1 \omega _1  + I_2 \omega _2  - 1 + \frac{3}
{8}I_2^2  - \frac{{4\omega _1  + 5}}
{{32\omega _1^2 \left( {\omega _1  + 1} \right)}}I_1^2  +  \hfill \\
   + \left[ { \pm \frac{{3\sqrt 2 }}
{2} - \frac{1} {{\omega _1 \left( {\omega _1  + 1} \right)}}}
\right]\frac{{I_1 I_2 }}
{{8\omega _1 }} \hfill \\
\end{gathered}
\end{equation}

In the left well among terms depending on angles we leave only two
terms: the resonance term
\begin{equation}\label{2.4.18}
{\raise0.7ex\hbox{$1$} \!\mathord{\left/
 {\vphantom {1 2}}\right.\kern-\nulldelimiterspace}
\!\lower0.7ex\hbox{$2$}}\frac{{I_1 }} {{\omega _1 }}\left(
{\frac{{2I_2 }} {{\omega _2 }}} \right)^{{\raise0.7ex\hbox{$1$}
\!\mathord{\left/
 {\vphantom {1 2}}\right.\kern-\nulldelimiterspace}
\!\lower0.7ex\hbox{$2$}}} \cos \left( {2\varphi _1  - \varphi _2 }
\right)
\end{equation}
and "swinging" term
\begin{equation}\label{2.4.19}
\frac{{\sqrt 2 }} {{64\omega _1 }}\left[ {\frac{3} {{\omega _1  +
1}} + \frac{1} {{\omega _1 }}} \right]I_1 I_2 \cos \left(
{2\varphi _1  - 2\varphi _2 } \right),
\end{equation}
the direction $m$ of which is the nearest to the resonance
direction r. After that the direct application of the destruction
stochastic layer criterion leads to the value of critical energy
in the left well $ E_{cr}  \approx  - 0.51$ which is in well
agreement both with the result of numerical simulation and with
the predictions of the negative curvature criterion $( - 5/9)$ .
Immediate analysis of integrable part of Hamiltonian
(\ref{2.4.17}) shows, that in the right well the resonances are
absent at negative energies and the transition to large-scale
stochasticity is possible only with the reach of saddle energy in
the complete correspondence with the numerical results .

We can briefly formulate the main results relating to the
definition of critical energy of the transition from regular
motion to stochastic one concluding comparison of the results of
numerical simulation with the analytical estimates .

1. The critical energy of the transition to chaos consistents well
with the predictions of the negative curvature criterion for the
PES with unique minimum.

2. The critical energy of the transition to chaos either is equal
to the minimal energy on the zero curvature line or coincides with
the saddle energy for the PES with few local minima .

3 The mixed state is observed in the interval $ U_{\min } \left(
{K = 0} \right) < E < E_s$ for the potential with several local
minima.

4. The possibility of application of negative curvature criterion
(NCC) to the particular local minimum is determined by the
structure of the Poincare surface of section: for any local
minimum,

\subsection{ Birkhoff-Gustavson normal form}

The structure of the Poincare surfaces of section can be
reproduced not resorting to the numerical solution of the
equations of motion. For this, let us use the method of treating
non separable classical systems that was originally developed by
Birkhoff \cite{45} and later was extended by Gustavson \cite{46}.
The result, obtained by Birkhoff, is the following: if the given
Hamiltonian $H$ , which can be written as a formal power series
without constant or linear terms, and such that the quadratic
terms can be written as a sum of uncoupled harmonic oscillator
terms with incommensurable frequencies, then there is a canonical
transformation that transforms original Hamiltonian into a normal
form. The normal form is a power series in one-dimensional
uncoupled harmonic oscillator Hamiltonian. Birkhoff's method was
applied by Gustavson in order to obtain power series expressions
for isolating integrals, and to predict analytically the Poincare
surfaces of section for the Henon-Helies system. Since the treated
potential had commensurable frequencies, Gustavson had to modify
Birkhoff's method only somewhat.

We shall consider the procedure of the transformation to
Birkhoff's normal form for Hamiltonian which is a power series in
coordinates $u$ and momenta $v$ \cite{47}
\begin{equation}\label{2.5.1}
H\left( {\vec u,\vec v} \right) = H^{\left( 2 \right)} \left(
{\vec u,\vec v} \right) + H^{\left( 3 \right)} \left( {\vec u,\vec
v} \right) + ...
\end{equation}
where $H^{(S)} $ has the form of a homogeneous polynomial of
degree $s$:
\[H^{\left( s \right)} \left( {\vec u,\vec v} \right) = \sum\limits_{\left| i \right|
 + \left| j \right| = s} {a_{ij} u^i v^j s = 2,3...}\]
\begin{equation}\label{2.5.2}
u^i  = u_1^{i_1 } u_2^{i_2 } ...u_N^{i_N } \quad \left| i \right|
= i_1  + i_2  + ... + i_N
\end{equation}
For systems in which $H^{(2)} $ is positive definite, then there
exists a canonical transformation $(\vec u,\vec v) \to (\vec
q,\vec p)$ which transforms $H^{(2)} $ into the form
\begin{equation}\label{2.5.3}
H^{\left( 2 \right)} \left( {\vec q,\vec p} \right) =
\sum\limits_{k = 1}^N {\frac{1} {2}\omega _k \left( {q_k^2  +
p_k^2 } \right)}
\end{equation}
Let $H(\vec q,\vec p)$ be Hamiltonian with $H^{(2)} (\vec q,\vec
p)$ as given in eq. (\ref{2.5.3}). Then we say that $H(\vec q,\vec
p)$ is in a normal form, if
\begin{equation}\label{2.5.4}
DH\left( {\vec q,\vec p} \right) = 0
\end{equation}
where
\begin{equation}\label{2.5.5}
 D =  - \sum\limits_k {\omega _k \left( {q_k \frac{\partial }
{{\partial p_k }} - p_k \frac{\partial } {{\partial q_k }}}
\right)}
\end{equation}

This is equivalent to the requiring that the Poisson bracket of
$H^{(2)} $ with $H$ vanish, since $D$ is given by
\begin{equation}\label{2.5.6}
D =  - \left[ {H^{\left( 2 \right)} ,} \right]
\end{equation}

A power series Hamiltonian can be transformed to the normal form
by a sequence of canonical transformations, where each one reduces
the non normalized term of the lowest degree to the normal form.
The generating function, necessary for the performance of
individual transformation, is defined by
\begin{equation}\label{2.5.7}
 F\left( {\vec P,\vec q} \right) = \sum\limits_k {P_k q_k  + W^{\left( s \right)} \left( {\vec P,\vec q} \right)}
\end{equation}
The connection between old $(\vec p,\vec q)$ and new $(\vec P,\vec
Q)$ canonical variables is
\begin{equation}\label{2.5.8}
 Q_k  = q_k  + \frac{{\partial W^{\left( s \right)} }}
{{\partial P_k }},\quad p_k  = P_k  + \frac{{\partial W^{\left( s
\right)} }} {{\partial q_k }},\quad H\left( {\vec p,\vec q}
\right) = \Gamma \left( {\vec P,\vec Q} \right)
\end{equation}
where $\Gamma (\vec P,\vec Q)$ is the Hamiltonian in the new
variables. If we expand $H$ and $\Gamma $ in a Taylor series about
$\vec P$ and $\vec q$ , and then collect and equate all terms of
equal degree, the following equation for $W^{(S)} $ is obtained
\begin{equation}\label{2.5.9}
 DW^{\left( s \right)} \left( {P,q} \right) = \Gamma ^{\left( s \right)} \left( {P,q} \right) - H^{\left( s \right)} \left( {P,q} \right)
\end{equation}
In order to solve eq. (\ref{2.5.9}) for $W^{(S)} $ , we make a
transformation to variables in which $D$ is diagonal:
\begin{equation}\label{2.5.10}
\begin{gathered}
  P_k  = 2^{ - {\raise0.7ex\hbox{$1$} \!\mathord{\left/
 {\vphantom {1 2}}\right.\kern-\nulldelimiterspace}
\!\lower0.7ex\hbox{$2$}}} \left( {\eta _k  + i\xi _k } \right) \hfill \\
  q_k  = i2^{{\raise0.7ex\hbox{$1$} \!\mathord{\left/
 {\vphantom {1 2}}\right.\kern-\nulldelimiterspace}
\!\lower0.7ex\hbox{$2$}}} \left( {\eta _k  - i\xi _k } \right) \hfill \\
\end{gathered}
\end{equation}
Under this transformation $ D \to \tilde D$ , where
\begin{equation}\label{2.5.11}
\tilde D = i\sum\limits_k {\omega _k \left( {\xi _k \frac{\partial
} {{\partial \xi _k }} - \eta _k \frac{\partial } {{\partial \eta
_k }}} \right)}
\end{equation}
It follows that functions of the form
\begin{equation}\label{2.5.12}
  \Phi _{l_1 l_2 m_1 m_2 }  \equiv \eta _1^{l_1 } \eta _2^{l_2 } \xi _1^{m_1 } \xi _2^{m_2 }
\end{equation}
are the eigenfunction of $ \tilde D$ , i.e.
\begin{equation}\label{2.5.13}
\tilde D\Phi _{l_1 l_2 m_1 m_2 }  = \left[ {i\sum\limits_k {\omega
_k \left( {m_k  - l_k } \right)} } \right]\Phi _{l_1 l_2 m_1 m_2 }
\end{equation}
Consequently,
\begin{equation}\label{2.5.14}
 \tilde D^{ - 1} \Phi _{l_1 l_2 m_1 m_2 }  = \left[ {i\sum\limits_k {\omega _k \left( {m_k  - l_k } \right)} } \right]^{ - 1} \Phi _{l_1 l_2 m_1 m_2 }
\end{equation}
Now the equation (\ref{2.5.9}) may  be solved for $\tilde W^{(S)}
$
\begin{equation}\label{2.5.15}
\tilde W^{\left( S \right)}  = \tilde D^{ - 1} \left( {\tilde
\Gamma ^{\left( S \right)}  - \tilde H^{\left( S \right)} }
\right)
\end{equation}
Here $ \tilde H^{\left( S \right)}$ is a known function. However,
so far, $\tilde \Gamma ^{\left( S \right)} $ has been unspecified.
Now  $\tilde \Gamma ^{(S)} $ is  determined from the requirement
that $ \tilde W^{\left( S \right)}$ is finite. Clearly, $\tilde
\Gamma ^{(S)} $ must be chosen so that to exactly cancel any terms
in $ \tilde H^{\left( S \right)} $ , which would give a vanishing
dominator in eq.(\ref{2.5.14}). So long as the frequencies are
incommensurable, the only terms that must appear in $\tilde \Gamma
^{\left( S \right)} $ are those for which $m_k  = l_k $ for all
$k$ ; those terms are
\begin{equation}\label{2.5.16}
 \left( {i\eta _1 \xi _1 } \right)^{m_1 } \left( {i\eta _2 \xi _2 } \right)^{m_2 }  = \left[ {\frac{1}
{2}\left( {P_1^2  + q_1^2 } \right)} \right]^{m_1 } \left[
{\frac{1} {2}\left( {P_2^2  + q_2^2 } \right)} \right]^{m_2 }
\end{equation}
Such terms are called null space terms, and the remaining terms
are called range space terms. Therefore, if $ \tilde H^{\left( S
\right)}$ is separated into null space terms $ \tilde N^{\left( S
\right)}$ and range space terms $ \tilde R^{\left( S \right)}$ ,
\begin{equation}\label{2.5.17}
 \tilde H^{(S)}  = \tilde N^{(S)}  + \tilde R^{(S)}
\end{equation}
and if we require $\tilde \Gamma ^{(S)} $ to cancel the null space
term in $ \tilde H^{\left( s \right)}$ , eq(\ref{2.5.15}) results
in
\begin{equation}\label{2.5.18}
 \begin{gathered}
  \tilde \Gamma ^{\left( S \right)}  = \tilde N^{\left( S \right)}  \hfill \\
  \tilde W^{\left( S \right)}  = D^{ - 1} \tilde R^{\left( S \right)}  \hfill \\
\end{gathered}
\end{equation}

In order to make this solution the unique one, it is sufficient to
require the generating function containing no null space terms.

If the transformation to Birkhoff's normal form is made, the
harmonic oscillator terms can be transformed to action variables
via the canonical transformation (\ref{2.4.10}).

We cite as an example the normal form (up to $s = 6$ ) for the
Hamiltonian (\ref{2.4.7}) at $a = 2$ in the neighbourhood of right
minimum
\begin{equation}\label{2.5.19}
  \begin{gathered}
  H\left( {I_1 ,I_2 } \right) = 2.613I_1  + 2I_2  - 0.219I_1 I_2  - 0.017I_1^2  - 0.375I_2^2  -  \hfill \\
   - 0.005I_1^3  - 0.028I_1^2 I_2  - 0.122I_1 I_2^2  - 0.133I_2^3  \hfill \\
\end{gathered}
\end{equation}

The commensurability of frequencies results into the extension of
the set of null space terms. As a matter of fact
\begin{equation}\label{2.5.20}
\tilde D\eta _1^{m_1 } \xi _2^{m_2 }  = i\left( {m_1 \omega _1  -
m_2 \omega _2 } \right)\eta _1^{m_1 } \xi _2^{m_2 }
\end{equation}
and at the realization of condition
\begin{equation}\label{2.5.21}
\left( {m_1 \omega _1  - m_2 \omega _2 } \right) = 0
\end{equation}
$ \tilde D^{ - 1} \eta _1^{m_1 } \xi _2^{m_2 } $ would diverge. In
order to avoid this occurrence, $ \tilde \Gamma ^{\left( S
\right)} $ must be chosen to cancel any such additional terms in $
\tilde H^{\left( S \right)}$ . Aside from this change, the
procedure is identical to that for the incommensurable case.

The reduction of Hamiltonian to the normal form solves the
question about the construction of full set of approximate
integrals of motion. The latter can be found by transformation of
the variables of action to initial variables. The solution of
equations
\begin{equation}\label{2.5.22}
\begin{gathered}
  H\left( {p_x ,p_y ,x,y} \right) = E \hfill \\
  I\left( {p_x ,p_y ,x,y} \right) = I_0  \hfill \\
  x = const \hfill \\
\end{gathered}
\end{equation}
allows to find the set of intersections of phase trajectory with
selected plane ($x = const$ ) and by doing so to reconstruct the
structure of the Poincare surfaces of section.

The Poincare surfaces of section for the quadrupole oscillations
of nuclei $Kr^{74} $ , which are constructed in such a way are
shown in Fig.8. The Hamiltonian, describing these oscillations, up
to the terms of the sixth degree with the respect to deformation
is the following
\begin{equation}\label{2.5.23}
\begin{gathered}
  H = \frac{1}
{2}\left( {p_x^2  + p_y^2 } \right) + U\left( {x,y} \right) \hfill \\
  U\left( {x,y} \right) = \frac{a}
{2}\left( {x^2  + y^2 } \right) + b\left( {x^2 y - \frac{1}
{3}y^3 } \right) + c\left( {x^2  + y^2 } \right)^2  +  \hfill \\
  d\left( {x^2 y - \frac{1}
{3}y^3 } \right)\left( {x^2  + y^2 } \right) + e\left( {x^2 y -
\frac{1}
{3}y^3 } \right)^2  + f\left( {x^2  + y^2 } \right)^3  \hfill \\
\end{gathered}
\end{equation}
\begin{figure}
\centering
\includegraphics[height=16cm]{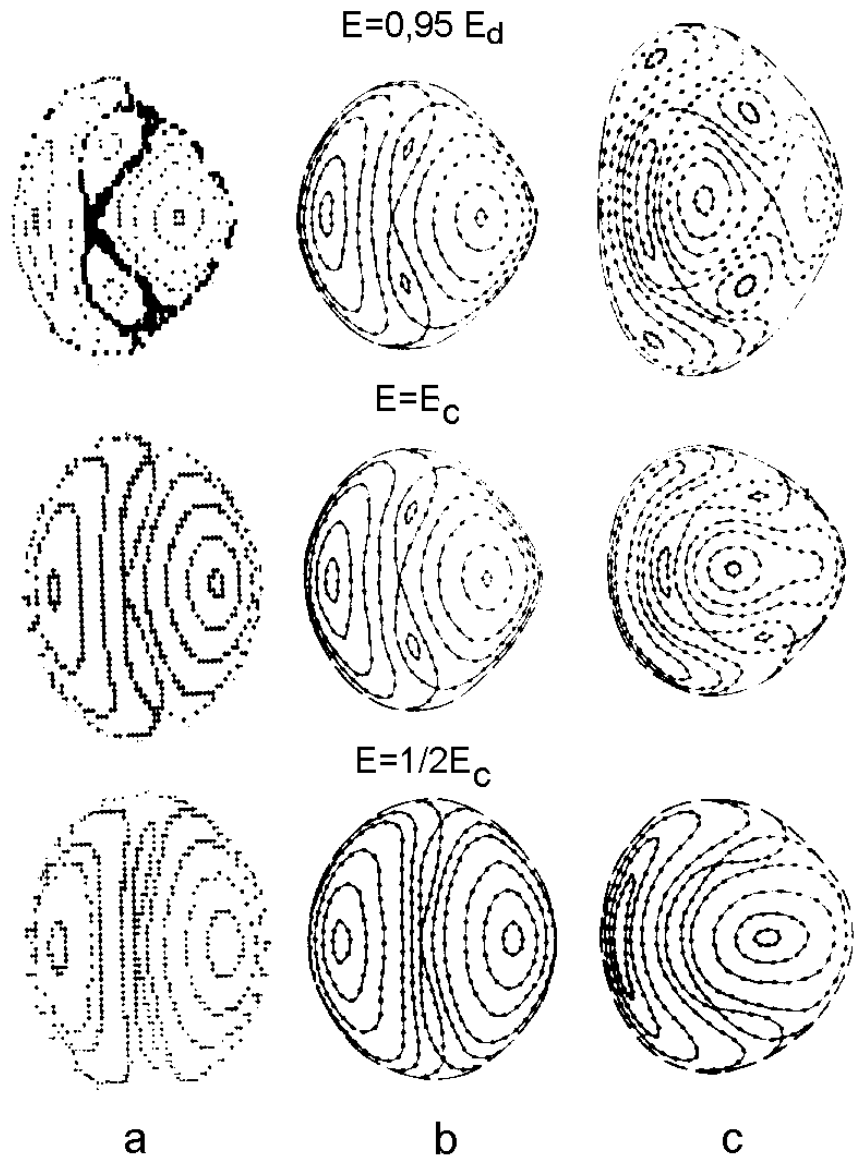}
\caption{The Poincare surfaces of section for different values of
energy for the central minimum of $Kr^{74} $ . a) obtained by the
numerical integration of equations of motion; b) obtained with the
help of normal forms; c) for the equivalent potential of
quadrupole oscillations of the fourth degree. } \label{fig.8}
\end{figure}

The parameters determining the dynamics of the particular nucleus,
is calculated for the isotopes of Krypton in paper \cite{29}. The
Poincare surfaces of section for the equivalent quartic potential,
the parameters of which are selected from the condition of
coincidence of the situation of critical points and the values of
energy in them are shown in the same figure.

\subsection{ Regularity-chaos-regularity transition}

The transition regularity-chaos for nonintegrable low-dimensional
Hamiltonian, which is going on as the energy or the amplitude of
external field increases, is a well-investigated process. The
critical energy of this transition, calculated within the
framework of different scenarios of stochastization, at least for
potentials with the simple geometry, is in the agreement with the
results of numerical simulation. For the systems with localized
region of instability (region of negative Gaussian curvature or
region of overlap of non-linear resonances) at further increasing
of energy, one would expect the return to the regular motion; and
the critical energy of this new transition chaos-regularity
$E_{cr2} $ will be determined by the top boundary of region of
instability. For the Hamiltonian of quadrupole oscillations
(\ref{2.1.8})
\begin{equation}\label{2.6.1}
E_{cr2}  = U\left( {x = 0,y_{cr2} } \right),y_{cr2}  =  - \frac{1}
{4}\left( {1 + \sqrt {1 - \frac{4} {W}} } \right)
\end{equation}
\begin{figure}[t]
\centering
\includegraphics[height=11cm]{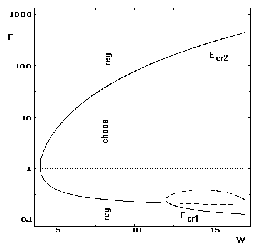}
\caption{Figure represents "the phase diagram" that allows to
determine energetic intervals of regular and chaotic motions for
the fixed value of $W$ .  Recall that for the PES with $0 < W < 4$
at all energies the motion remains regular. } \label{fig.9}
\end{figure}
\begin{figure}
\centering
\includegraphics[height=6cm]{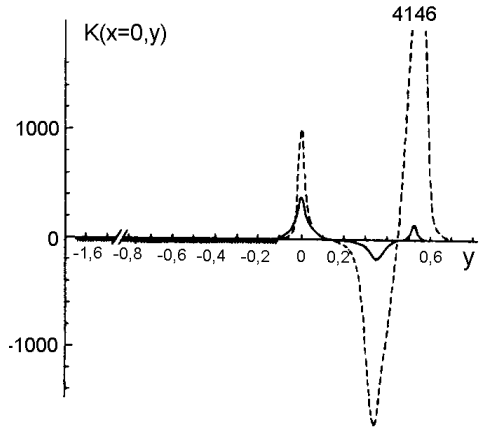}
\caption{The value of Gaussian curvature $K\left( {x = 0,y}
\right)$ of the PES of $Kr^{74} $ (solid line) and the value of
the equivalent potential of the fourth degree (dashed line). The
value of $K$ in the shaded range of $y$ are very small (less then
0.1). } \label{fig.10}
\end{figure}
\begin{figure}
\centering
\includegraphics[height=6cm]{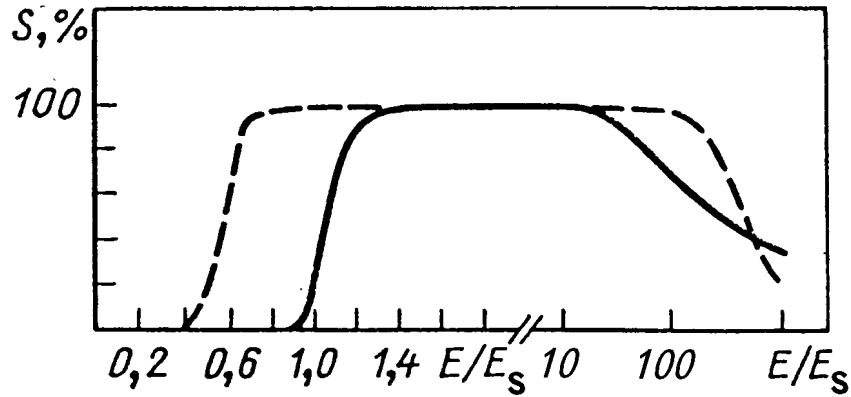}
\caption{ The part of phase space $S\% $ .with chaotic traectories
as a function of the energy ($E_S $ - the saddle energy). The
solid line - for potential of isotope $Kr^{74} $ and the dashed
line - for equivalent potential of the fourth degree. It is seen
that for the nucleus of Krypton the critical energy of the
transition chaos-regularity approximately twice exceeds the
critical energy for the equivalent potential. With increasing of
energy in both potentials the $100\% $ chaos appears. However, at
energies considerably exceeded the saddle energy, and for the
nucleus $Kr^{74} $ and for the equivalent potential the regular
character of motion is restoring. } \label{fig.11}
\end{figure}
Recall that for the PES with $0 < W < 4$ at all energies the
motion remains regular. Fig.9 represents "the phase diagram" that
allows to determine energetic intervals of regular and chaotic
motions for the fixed value of $W$ .

In the connection with the discussion of possibility of existence
of additional transition chaos-regularity, let us trace the change
of sign and absolute value of Gaussian curvature for the PES of
$Kr^{74} $ (\ref{2.5.23}) and the equivalent PES of the forth
degree (\ref{2.1.3}). The parameters of the equivalent PES are
fitted according to the condition of coincidence of the position
of extremuma of these two potentials and values of energy in them.
The value of Gaussian curvature $K(x = 0,y)$ of the PES of
$Kr^{74} $ and the value of the  equivalent potential of the
fourth degree are represented in Fig.10. We see, that the region
of negative curvature of the PES of the equivalent potential
occupies considerably larger region of space and has one order of
value larger at $y > 0$ than for the PES of $Kr^{74} $ . The
measure of divergence of the classical trajectories, which leads
to the rise of the stochastic properties in the system, is
determined by the size of the region and absolute value of
negative Gaussian curvature. This circumstance allows us to
understand qualitatively the reason of transition to chaos in
nucleus $Kr^{74} $ at comparatively higher energies, than in the
equivalent potential: the factors which determine the chaotic
character of motion in nucleus $Kr^{74} $ are essentially
suppressed. Comparatively small region of space, where the
negative Gaussian curvature of the PES of isotope $Kr^{74} $ is
localized, determines the character of motion at energies
essentially exceeded the saddle energy. The part of chaotic
trajectories $S,\% $ in relation to all considered trajectories as
a function of energy for the motion in deformation potential of
isotope $Kr^{74} $ and equivalent potential of the fourth degree
is represented in Fig. 11. It is seen that for the nucleus of
Krypton the critical energy of the transition chaos-regularity
approximately twice exceeds the critical energy for the equivalent
potential. In both potentials the $100\% $ chaos appears with
increasing of energy. However, at energies considerably exceeded
the saddle energy, and for the nucleus $Kr^{74} $ and for the
equivalent potential the regular character of motion is restoring.
This new transition chaos-regularity is illustrated by the
Poincare surfaces of section at the super higher energies, which
are represented in Fig.12.
\begin{figure}
\centering
\includegraphics[height=14cm]{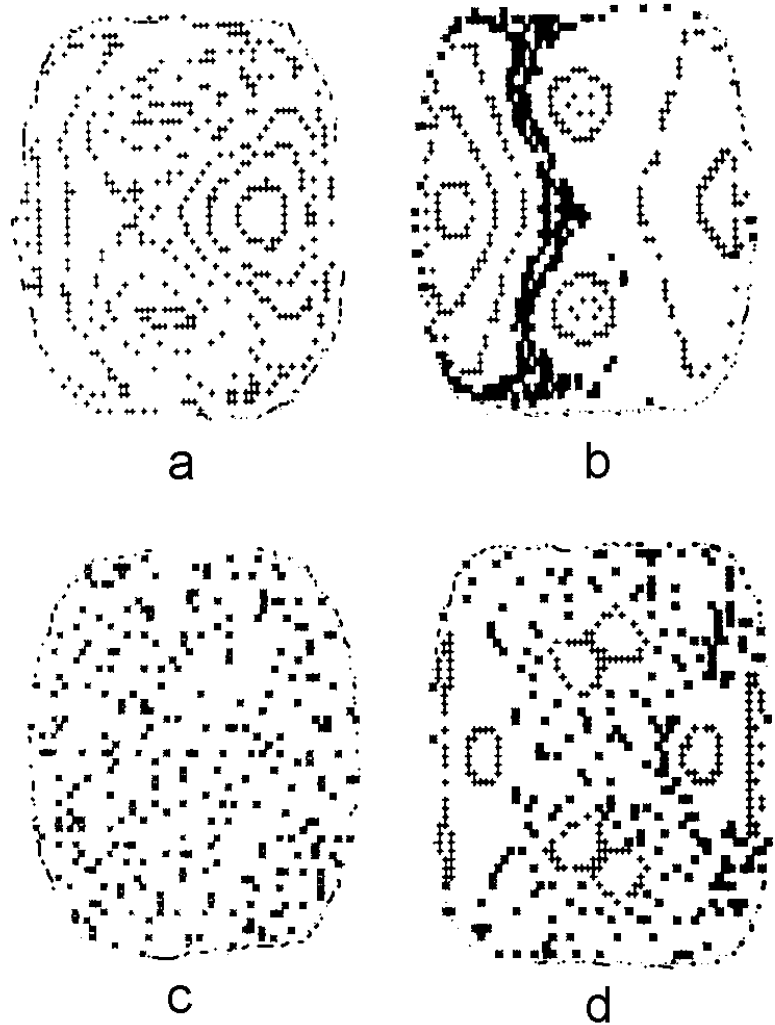}
\caption{ The Poincare surfaces of sections illustrating the
restoration regular character of motion at energies considerably
exceeded the saddle energy for the nucleus $Kr^{74} $ (b,d) and
for the equivalent potential (a,c). The figures (c,d) at the
middle energy: $E = 100E_S $ , and (a,b) - at high energy: $E =
1000E_S $ . } \label{fig.12}
\end{figure}

An earlier appearing of regular motion, at high energies for the
quadrupole oscillations of nuclear surface of $Kr^{74} $ comparing
with the oscillations in equivalent potential, as with low
energies, is explained by smaller absolute value of Gaussian
curvature and larger degree of its localization. Underline, that
similar reconstruction of regular motion at high energies must
occur for any potential with localized region of instability. In
particular, it occurs for isotope of $Kr^{76} $ . For isotopes
$Kr^{78,80} $ the negative curvature is not spatially localized
and that is why the regular character of motion is not
reconstructed at energies accessible to numerical calculation.

In conclusion of this section we would like to note the similarity
in structure of phase space of considered two-dimensional
autonomous Hamiltonian system with the compact region of negative
Gaussian curvature and one-dimensional system with periodic
perturbation \cite{48}.

The behavior of the width of the resonances, $ \bar W_k \equiv
\frac{1} {2}\left( {W_{k + 1}  + W_k } \right)$ , and the
distances between them, $ \Delta I_k  \equiv \left| {I_{k + 1}  -
I_k } \right|$ , as a function of the resonance number is the
simplest when the satisfaction of resonance overlap condition
(2.3.20) for number $k_1 $ (at a fixed level of the external
perturbation) guarantees that this condition holds for arbitrary
$k > k_1 $ . This is precisely the situation which prevails in the
extensively studied systems of a $1D$ Coulomb potential \cite{49}
and a square well \cite{50} subjected in each case to a
monochromatic perturbation. In the former case we have $ \bar W_k
\approx k^{{\raise0.7ex\hbox{$1$} \!\mathord{\left/
 {\vphantom {1 6}}\right.\kern-\nulldelimiterspace}
\!\lower0.7ex\hbox{$6$}}}$ and $ \Delta I_k  \approx k^{ -
{\raise0.7ex\hbox{$2$} \!\mathord{\left/
 {\vphantom {2 3}}\right.\kern-\nulldelimiterspace}
\!\lower0.7ex\hbox{$3$}}}$ , while in the latter we have $ \bar
W_k \approx k^{ - 1}$ and $ \Delta I_k  \approx \left[ {k\left( {k
+ 1} \right)} \right]^{ - 1}$ .
\begin{figure}[t]
\centering
\includegraphics[height=7.5cm]{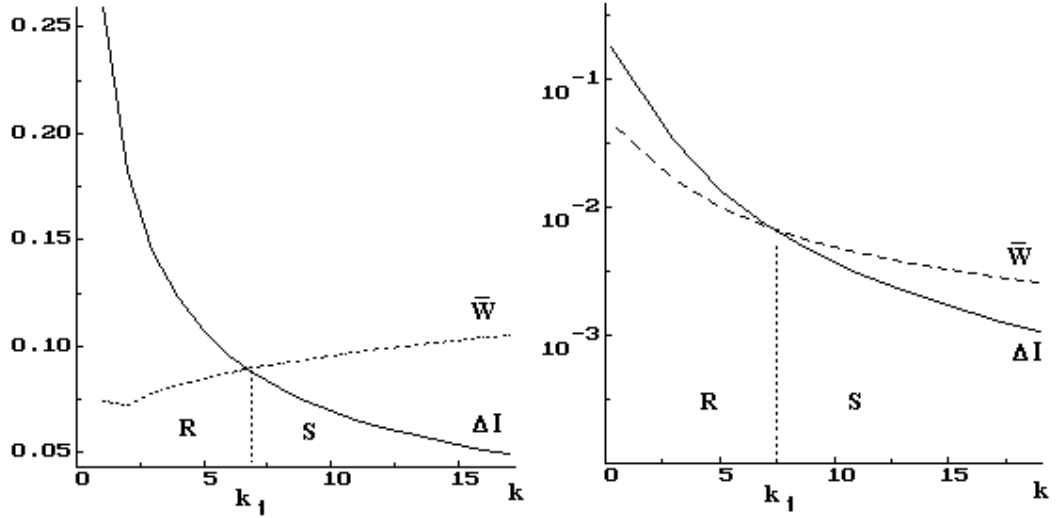}
\caption{ The resonant spacings $\Delta I_k $ and the mean widths
$ \bar W_k $ as functions of the resonance numbers $k$ . On the
left: for one-dimensional Coulomb, on the right: for square-well
potential. The critical point $k_1 $ separates regular range
$\left( R \right)$ from the chaotic one $\left( C \right)$ . }
\label{fig.13}
\end{figure}

As can be seen from Figure 13 there is a regularity-chaos
transition (we will call this a "normal" transition) for both $1D$
Coulomb problem and a square well,since there exists a unique
point $k_1 $ such that at $k > k_1 $ the condition $ \bar W_k  >
\Delta I_k$ always holds. The motion is therefore chaotic.
However, as the behavior of the widths of the resonances and of
the distances between them as a function of the resonance number
becomes more complex, we can allow the appearance of an additional
intersection point and thus a new transition: a chaos-regularity
transition, which we will call "anomalous". This is also the
exotic possibility of the intermittent occurrence of a regular and
chaotic regions in the phase space.

We demonstrate that an anomalous chaos-regularity transition
occurs in a simple Hamiltonian system: an anharmonic oscillator,
subjected to a monochromatic perturbation \cite{48}. The dynamics
of such system is generated by the Hamiltonian
\begin{equation}\label{2.6.2}
H\left( {p,x,t} \right) = H_0 \left( {p,x} \right) + Fx\cos \Omega
t
\end{equation}
with the unperturbed Hamiltonian is
\begin{equation}\label{2.6.3}
H_0 \left( {p,x} \right) = \frac{{p^2 }} {{2m}} + Ax^n  = E\quad
\left( {n = 2l,l > 1} \right)
\end{equation}
Considered system fills a gap between two extremely important
physical models: the harmonic oscillator ($n = 2$ ) and square
well ($n = \infty$ ). In terms of action-angle variables $(I,q)$ ,
the Hamiltonian $H_0 (p,x)$ becomes
\begin{equation}\label{2.6.4}
H_0 \left( I \right) = \left( {\frac{{2\pi }} {{\alpha G\left( n
\right)}}I} \right)^\alpha
\end{equation}
where
\begin{equation}\label{2.6.5}
 G\left( n \right) = \frac{{2\sqrt {2\pi m} \Gamma \left( {1 + \frac{1}
{n}} \right)}} {{A^{{\raise0.7ex\hbox{$1$} \!\mathord{\left/
 {\vphantom {1 n}}\right.\kern-\nulldelimiterspace}
\!\lower0.7ex\hbox{$n$}}} \Gamma \left( {\frac{1} {2} + \frac{1}
{n}} \right)}},\alpha  = \frac{{2n}} {{n + 2}}
\end{equation}
The resonant values of the action $I_k $ that can be found from
the condition $ k\omega \left( {I_k } \right) = \Omega ,\quad
\omega \left( I \right) = \frac{{dH_0 }} {{dI}}$ are
\begin{equation}\label{2.6.6}
 I_k  = \alpha \left( {\frac{{G\left( n \right)}}
{{2\pi }}} \right)^{2n\beta } \left( {\frac{\Omega } {k}}
\right)^{2n{\raise0.7ex\hbox{$\alpha $} \!\mathord{\left/
 {\vphantom {\alpha  \beta }}\right.\kern-\nulldelimiterspace}
\!\lower0.7ex\hbox{$\beta $}}} ,\beta  = \frac{1} {{n - 2}}
\end{equation}
A classical analysis, based on the resonance-overlap criterion,
leads to the following expression for the critical amplitude of
the external perturbation
\begin{equation}\label{2.6.7}
 F_k^{cr}  = 2^{\left( {2 - 3n} \right)\beta } \frac{1}
{{4n}}\frac{{\alpha ^2 }} {\beta }\frac{1} {{x_k }}\left(
{\frac{{G\left( n \right)}} {\pi }} \right)^{2n\beta } \Omega
^{2n\beta } k^{4\beta } \left[ {k^{\left( {n + 2} \right)\beta } -
\left( {k + 1} \right)^{\left( {n + 2} \right)\beta } } \right]^2
\end{equation}
where $x_k $ is a Fourier component of the coordinate $x(I,q)$ .
Expression (\ref{2.6.7}) solves the problem of reconstructing the
structure of the phase space for arbitrary values of the
parameters.
\begin{figure}[t]
\centering
\includegraphics[height=7.5cm]{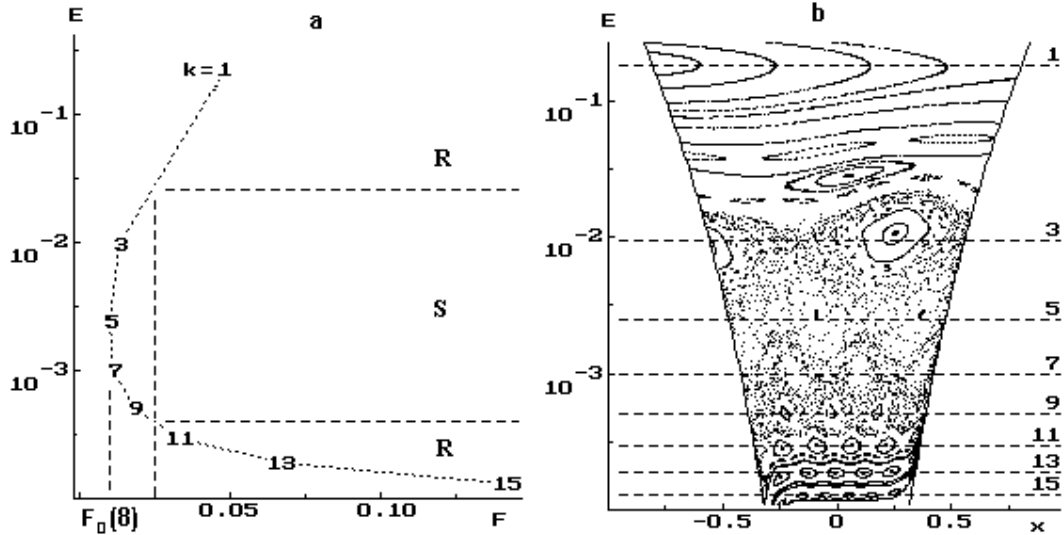}
\caption{The "phase diagram" - Figure (a) - the resonance energies
$E_k $ ($k$ - the resonance number) against the critical values of
the external perturbation for Hamiltonian (\ref{2.6.3}). The upper
part of curve $\left( {k = 1 - 5} \right)$ corresponds to a normal
transition regularity-chaos but the lower part of curve $\left( {k
= 5 - 15} \right)$ corresponds to an anomalous transition
chaos-regularity. The snap-shot of $E\left( x \right)$ at the
right in Figure (b) confirms that an anomalous chaos-regularity
transition occurs. We can clearly see isolated nonlinear
resonances that persist at large values of $k$ , and near which
the motion remains regular.} \label{fig.14}
\end{figure}

The "phase diagram" in Fig. 14a can be used to determine, at the
fixed level of the external perturbation, the energy intervals of
regular and chaotic motion. The snap-shot of $E(x)$ at the right
in Fig. 14b confirms that an anomalous chaos-regularity transition
occurs. We can clearly see isolated nonlinear resonances which
persist at large values of $k$ , and near which the motion remains
regular. The reason for this anomaly is explained by Fig.15. The
plots of the resonance widths and of the distances between
resonances in this Figure demonstrate that there are two
intersection points $k = k_1 $ and $k = k_2 $ , rather than one.
Consequently, there is an anomalous chaos-regularity transition.
\begin{figure}[t]
\centering
\includegraphics[height=10cm]{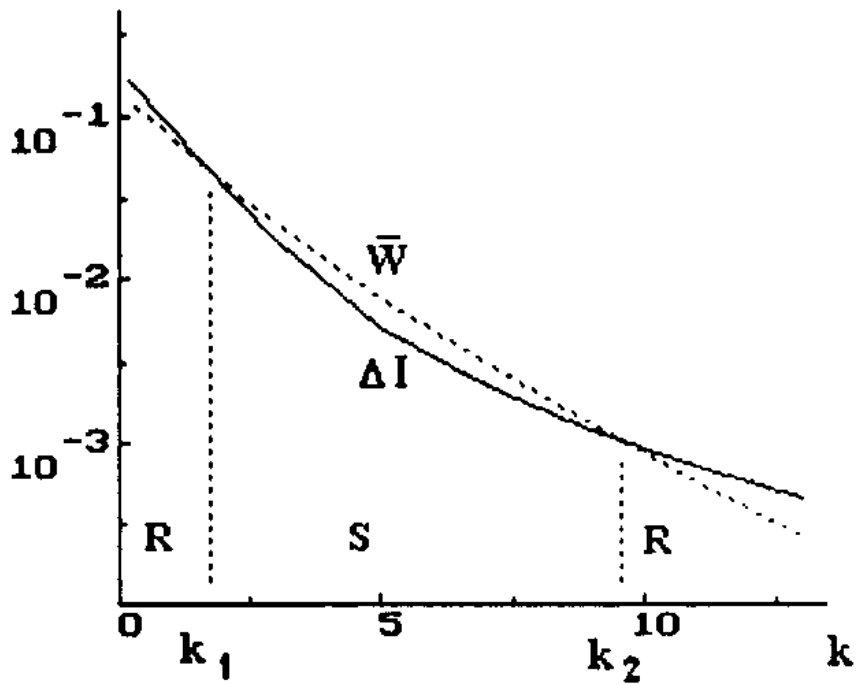}
\caption{The resonant spacings  $ \Delta I_k $ and the mean widths
$ \bar W_k $ as functions of the resonance numbers $k$ for
potential (\ref{2.6.3}) with $n = 8$ . One can see two critical
points $k_1 $ and $k_2 $ . The second critical point corresponds
to an anomalous chaos regularity transition. } \label{fig.15}
\end{figure}

Thus, one can observe the transition regularity-chaos- regularity
just as in the case of $2D$ autonomous Hamiltonian system, so in
the case of $1D$ system with periodic perturbation. The reason of
the additional transition in both cases is common: localized
region of instability. In the first case this reason is a
localized region of negative Gaussian curvature, in the second one
this reason is a localized region of overlap of non-linear
resonances.

\subsection{ Chaotic regimes in reactions with heavy ions}

Outlined in the previous sections, the common conception,
concerning ossible stochastization of quadrupole nuclear
oscillations of high amplitude, is confirmed by the direct
observations of chaotic regimes at mathematical simulation of
reactions with heavy ions \cite{8}.

The time-dependent Hartree-Fock (TDHF) theory \cite{27,51}
constitutes a well-defined starting point for the study of such
processes. The TDHF equations can be obtained from the variation
of the many-body action $S$ ,
\begin{equation}\label{2.7.1}
  S = \int\limits_{t_1 }^{t_2 } {dt\left\langle {\Psi \left( t \right)\left| {i\frac{\partial }
{{\partial t}} - H} \right|\Psi \left( t \right)} \right\rangle }
\end{equation}
In this expression $H$ is the many-body Hamiltonian and the $A$
-nucleon wave function $\Psi (t)$ is chosen to be of determinantal
form, constructed from the time-dependent single-particle states
$\psi _\lambda  \left( {\vec r,t} \right)$
\begin{equation}\label{2.7.2}
\Psi \left( {\vec r_1 ,\vec r_2 ...\vec r_A ;t} \right) = \frac{1}
{{\sqrt {A!} }}\det \left\| {\psi _\lambda  \left( {\vec r,t}
\right)} \right\|
\end{equation}
The variation of eq.(\ref{2.7.1}) is an independent variation with
the respect to the single-particle states $\psi _\lambda  $ and
$\psi _\lambda  ^ *  $ and yields the equations of motion,
\begin{equation}\label{2.7.3}
i\frac{\partial } {{\partial t}}\psi _\lambda  \left( {\vec r,t}
\right) = \frac{{\delta \left\langle H \right\rangle }} {{\delta
\psi _\lambda ^ *  \left( {\vec r,t} \right)}} \equiv h\left(
{\vec r,t} \right)\psi _\lambda  \left( {\vec r,t} \right)
\end{equation}
and a similar equation for $ \psi _\lambda ^ *  \left( {\vec r,t}
\right)$ .

The classical nature of these equations can be put into a better
respective via the definition of classical field coordinates $\phi
_\lambda  \left( {\vec r,t} \right)$ , and conjugate momenta $ \pi
_\lambda  \left( {\vec r,t} \right)$
\begin{equation}\label{2.7.4}
  \begin{gathered}
  \phi _\lambda   = {\raise0.7ex\hbox{${\left( {\psi _\lambda
   + \psi _\lambda ^ *  } \right)}$} \!\mathord{\left/
 {\vphantom {{\left( {\psi _\lambda   + \psi _\lambda ^ *  }
 \right)} {\sqrt 2 }}}\right.\kern-\nulldelimiterspace}
\!\lower0.7ex\hbox{${\sqrt 2 }$}} \hfill \\
  \pi _\lambda   = {\raise0.7ex\hbox{${\left( {\psi _\lambda
   - \psi _\lambda ^ *  } \right)}$} \!\mathord{\left/
 {\vphantom {{\left( {\psi _\lambda   - \psi _\lambda ^ *  }
 \right)} {\sqrt 2 }}}\right.\kern-\nulldelimiterspace}
\!\lower0.7ex\hbox{${\sqrt 2 }$}} \hfill \\
\end{gathered}
\end{equation}
Then the result is the Hamiltonian's equations
\begin{equation}\label{2.7.5}
\begin{gathered}
  \frac{{d\phi _\lambda  \left( {\vec r,t} \right)}}
{{dt}} = \frac{\begin{gathered}
   \hfill \\
  \delta \left\langle H \right\rangle  \hfill \\
\end{gathered} }
{{\delta \pi _\lambda  \left( {\vec r,t} \right)}} \hfill \\
  \frac{{d\pi _\lambda  \left( {\vec r,t} \right)}}
{{dt}} =  - \frac{{\delta \left\langle H \right\rangle }}
{{\delta \phi _\lambda  \left( {\vec r,t} \right)}} \hfill \\
\end{gathered}
\end{equation}
The TDHF equation (\ref{2.7.3}) and its complex conjugate are
solved on a three-dimensional space time lattice \cite{52} with
initial wave functions of the form
\begin{equation}\label{2.7.6}
 \begin{gathered}
  \lim \phi _\lambda  \left( {\vec r,t} \right) \to \sqrt 2 \cos \left( {\vec k_\lambda  \vec r - \varepsilon _\lambda  t} \right)\chi _\lambda  \left( {\vec r} \right) \hfill \\
  \lim \pi _\lambda  \left( {\vec r,t} \right) \to \sqrt 2 \sin \left( {\vec k_\lambda  \vec r - \varepsilon _\lambda  t} \right)\chi _\lambda  \left( {\vec r} \right) \hfill \\
\end{gathered}
\end{equation}
where $\varepsilon _\lambda  $ is the solution of the static HF
equations $ h\chi _\lambda  \left( {\vec r} \right) = \varepsilon
_\lambda  \chi _\lambda  \left( {\vec r} \right)\lambda  = 1,...A$
and $\vec k_\lambda  $ is the parameter of the initial boost.
\begin{figure}
\centering
\includegraphics[height=8cm]{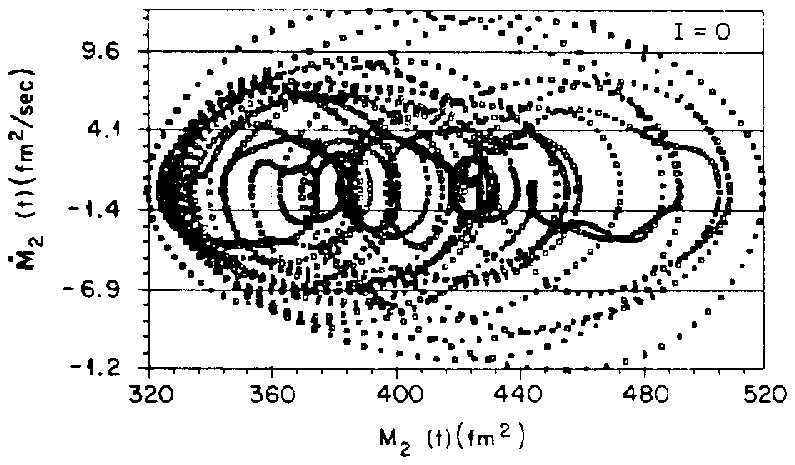}
\caption{Poincare phases space plot of $ \dot M_{LI} \left( t
\right)vsM_{LI} \left( t \right)$ for isoscalar quadrupole mode
for the $Mg^{24} $ system \cite{8}.} \label{fig.16}
\end{figure}

TDHF calculations for head-on collisions of $He^4  + C^{14}
,C^{12}  + C^{12} (0^ +  ),$ and $He^4  + Ne^{20} $ have been
performed by Umar et. all. \cite{8} at bombarding energies near
the Coulomb barrier. The results are interpreted in the terms of
their classical behavior. The initial energy and the separation of
the centers of the ions are the parameters labeling the initial
state. After the initial contact compound nuclear system ($O^{18}
$ or $Mg^{24} $ ) relaxes into a configuration, undergoing quasi
periodic or chaotic motion. The analysis of nuclear density
multipole moments $ \left\{ {M_{LI} \left( t \right),\dot M_{LI}
\left( t \right)} \right\}$ has been applied for classifying those
solutions Poincare sections. The definitions of the moments are as
follows
\begin{equation}\label{2.7.7}
 \begin{gathered}
  M_{LI} \left( t \right) = \int {d^3 rr^L Y_{LM} \left(
  {\hat r} \right)} \rho _I \left( {\vec r,t} \right) \hfill \\
  M_{LI} \left( \omega  \right) = \int {dt\exp \left( { - i\omega t}
   \right)M_{LI} } \left( t \right) \hfill \\
\end{gathered}
\end{equation}
where isoscalar ($I = 0$ ) and isovector ($I = 1$ ) densities,
\begin{equation}\label{2.7.8}
\rho _I \left( {\vec r,t} \right) = \left\{ {\begin{array}{*{20}c}
   {\rho _p \left( {\vec r,t} \right) + \rho _n \left( {\vec r,t} \right)\quad I = 0}  \\
   {\rho _p \left( {\vec r,t} \right) - \rho _n \left( {\vec r,t} \right)\quad I = 1}  \\
 \end{array} } \right.
\end{equation}
The proton $\rho _p $ and neutron $\rho _n $ densities in terms of
the field coordinates $\phi _l $ and momenta $\pi _l $ , are
\begin{equation}\label{2.7.9}
\rho _q \left( {\vec r,t} \right) = \frac{1}
{2}\sum\limits_\lambda  {\left[ {\left| {\pi _{\lambda ,q} \left(
{\vec r,t} \right)} \right|^2  + \left| {\phi _{\lambda ,q} \left(
{\vec r,t} \right)} \right|^2 } \right],q = p,n}
\end{equation}
\begin{figure}
\centering
\includegraphics[height=8cm]{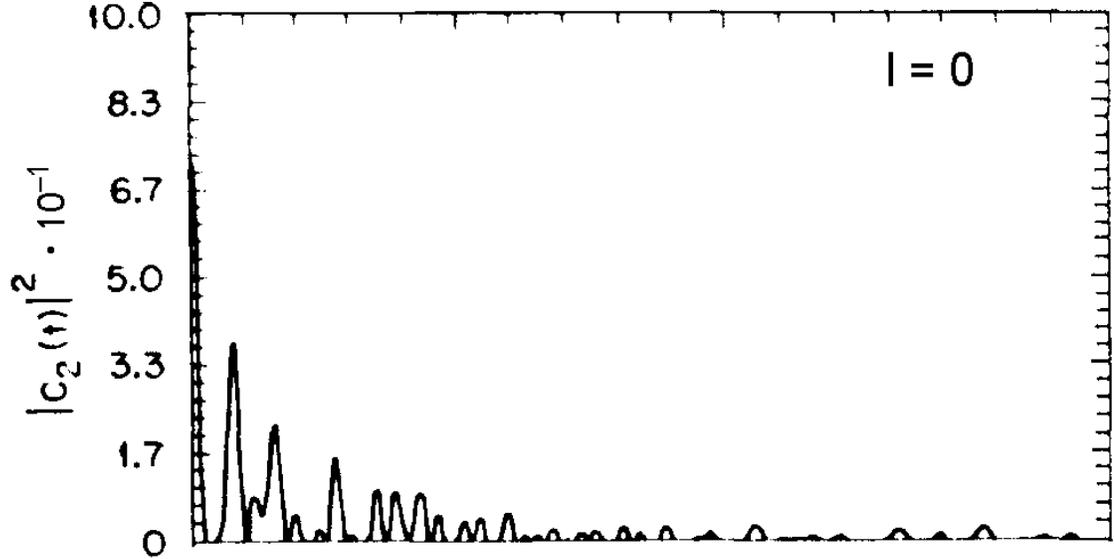}
\caption{The autocorrelation function $C_{LI} \left( t \right)$ in
units $Mev\left( {fm^L \;\sec } \right)^2 $ as a function of time
for the isoscalar quadrupole mode in the $Mg^{24} $ system
\cite{8}. } \label{fig.17}
\end{figure}
The isoscalar quadrupole mode ($L = 2,I = 0$ ) is shown in Fig.16
for the $Mg^{24} $ nuclear system seems to be filling most of the
available phase space in Poincare section $ \left\{ {M_{LI} \left(
t \right),\dot M_{LI} \left( t \right)} \right\}$ . The
corresponding autocorrelation function (Fig.17)

\begin{equation}\label{2.7.10}
C_{20} \left( t \right) = \int\limits_{ - \infty }^\infty
{\frac{{d\omega }} {{2\pi }}\exp \left( {i\omega t} \right)\left|
{M_{20} \left( \omega  \right)} \right|^2 }
\end{equation}
damps fast. All this favours the view that corresponding motion is
closer to be stochastic rather than quasiperiodic.

\end{document}